\documentclass[prb,twocolumn,showpacs,floatfix,amsmath,amssymb,superscriptaddress]{revtex4-1}
\usepackage{amsfonts}
\usepackage{stmaryrd}
\usepackage{bbm}
\usepackage{mathrsfs}
\usepackage{tipa}
\usepackage{amssymb}
\usepackage{txfonts}
\usepackage{graphicx}
\usepackage{dcolumn}
\usepackage{epstopdf}
\usepackage[colorlinks,linkcolor=blue,urlcolor=blue,citecolor=blue]{hyperref}
\usepackage{multirow}
\usepackage{subfigure}
\usepackage{url}
\usepackage{upgreek}
\usepackage{booktabs}

\begin{document}
\newcommand*{\cm}{cm$^{-1}$\,}
\title{Photo-induced nonequilibrium response in underdoped YBa$_2$Cu$_3$O$_{6+x}$ probed by time-resolved terahertz spectroscopy }

\author{S. J. Zhang}
\affiliation{International Center for Quantum Materials, School of Physics, Peking University, Beijing 100871, China}

\author{Z. X. Wang}
\affiliation{International Center for Quantum Materials, School of Physics, Peking University, Beijing 100871, China}

\author{H. Xiang}
\affiliation{School of Physics and Astronomy, Shanghai Jiao Tong University, Shanghai 200240, P. R. China}

\author{X. Yao}
\affiliation{School of Physics and Astronomy, Shanghai Jiao Tong University, Shanghai 200240, P. R. China}

\author{Q. M. Liu}
\affiliation{International Center for Quantum Materials, School of Physics, Peking University, Beijing 100871, China}

\author{L. Y. Shi}
\affiliation{International Center for Quantum Materials, School of Physics, Peking University, Beijing 100871, China}

\author{T. Lin}
\affiliation{International Center for Quantum Materials, School of Physics, Peking University, Beijing 100871, China}

\author{T. Dong}
\affiliation{International Center for Quantum Materials, School of Physics, Peking University, Beijing 100871, China}

\author{D. Wu}
\affiliation{International Center for Quantum Materials, School of Physics, Peking University, Beijing 100871, China}

\author{N. L. Wang}
\email{nlwang@pku.edu.cn}
\affiliation{International Center for Quantum Materials, School of Physics, Peking University, Beijing 100871, China}
\affiliation{Collaborative Innovation Center of Quantum Matter, Beijing, China}

\begin{abstract}
Intense laser pulses have recently emerged as a tool to tune between different orders in complex quantum materials. Among different light-induced phenomena, transient superconductivity far above the equilibrium transition temperature in cuprates is particularly attractive. Key to those experiments was the resonant pumping of specific phonon modes, which was believed to induce superconducting phase coherence by suppressing the competing orders or modifying the structure slightly. Here, we present a comprehensive study of photo-induced nonequilibrium response in underdoped YBa$_2$Cu$_3$O$_{6+x}$. We find that upon photo-excitations, Josephson plasma edge in superconducting state is initially removed accompanied by quasiparticle excitations, and subsequently reappears at frequency lower than the static plasma edge within short time. In normal state, an enhancement or weaker edge-like shape is indeed induced by pump pulses in the reflectance spectrum accompanied by simultaneous rises in both real and imaginary parts of conductivity. We compare the pump-induced effects between near- and mid-infrared excitations and exclude phonon pumping as a scenario for the photo-induced effects above. We further elaborate the transient responses in normal state are unlikely to be explained by photo-induced superconductivity.
\end{abstract}

\pacs{}

\maketitle

High-T$_c$ superconducting cuprates (HTSC) are highly anisotropic materials. The conducting CuO$_2$ layers are separated by different block layers, leading to less-conducting or insulator-like c-axis \emph{dc} and optical responses in normal state. The low frequency c-axis optical spectra of HTSC are dominated by infrared active phonons with very small contribution from free carriers. However, when the cuprate system goes into superconducting state, a very sharp plasma edge suddenly develops in the c-axis reflectivity spectrum. The plasma edge is caused by the condensed superfluid carriers via Josephson tunneling effect, which is referred to as Josephson plasmon edge (JPE) \cite{PhysRevLett.69.1455,Uchida1996,PhysRevLett.71.1645}. Manifestation of such c-axis JPE is taken as an optical evidence for the occurrence of superconductivity.

Ultrashort pulses now provide unique opportunities to manipulate different orders and physical properties in complex electronic materials. An unexpected finding is that intense ultrafast excitation can induce a Josephson plasma-like edge and 1/$\omega$-like dependence in imaginary part of conductivity in normal state of cuprates, which was taken as an indication of light-induced transient superconductivity. The effect was first observed notably 5 ps after excitation in a stripe-ordered cuprate La$_{1.8-x}$Eu$_{0.2}$Sr$_x$CuO$_4$ at 10 K whose T$_c$ is less than 2 K \cite{Fausti189}, then in underdoped YBa$_2$Cu$_3$O$_{6+x}$ (YBCO) at temperature even above room temperature \cite{Kaiser2014,hu2014optically}. In those measurements, the pump excitation wavelength is tuned to about 15 $\upmu$m ($\sim$ 20 THz) in mid-infrared (MIR) region which is assumed to be resonant with specific phonon modes of oxygens (e.g. in-plane Cu-O stretching mode or apical oxygen mode relative to CuO$_2$ planes). The resonant pumping of specific phonon mode is believed to be the key to light-induced superconductivity. The observations also motivated theoretical studies on transient superconductivity along the direction of the resonant phonon pumping \cite{Subedi2014,PhysRevLett.114.137001,PhysRevB.91.184506,PhysRevLett.118.087002,Okamoto2017,Bittner2019,Klein}.

Near-infrared (NIR) pump was also used to investigate the photoexcitation-induced effect in cuprates. Nicoletti et al. reported a surprisingly large effect on La$_{2-x}$Ba$_x$CuO$_4$ (x=0.115): a blue shift of JPE with even enhanced edge amplitude when T$<$T$_c$, and a light-induced edge when T$>$T$_c$ in the stripe ordered state whose amplitude is comparable with that seen in the equilibrium superconducting state \cite{Nicoletti2014,PhysRevB.91.174502}. Significant NIR pump-induced effect was recently also observed in time-resolved THz measurement on other compositions of La$_{1-x}$A$_x$CuO$_4$ (A=Ba, Sr) and electron-doped Pr$_{1-x}$LaCe$_x$CuO$_4$ \cite{PhysRevB.98.020506,PhysRevB.98.224507,PhysRevLett.121.267003,Cremin2019,PhysRevB.100.104507}. Those results are notably different from the above reports on La$_{1.8-x}$Eu$_{0.2}$Sr$_x$CuO$_4$ \cite{Fausti189} and YBa$_2$Cu$_3$O$_{6.5}$ \cite{Kaiser2014,hu2014optically}, where the relative change of reflectivity is less than 20$\%$ even after taking the penetration depth mismatch into account. Nonetheless, up to now there is no report about c-axis terahertz measurement on YBCO driven by NIR pulse excitations.

\begin{figure}[htbp]
  \centering
\includegraphics[width=8.7cm, trim=0 0 0 0,clip]{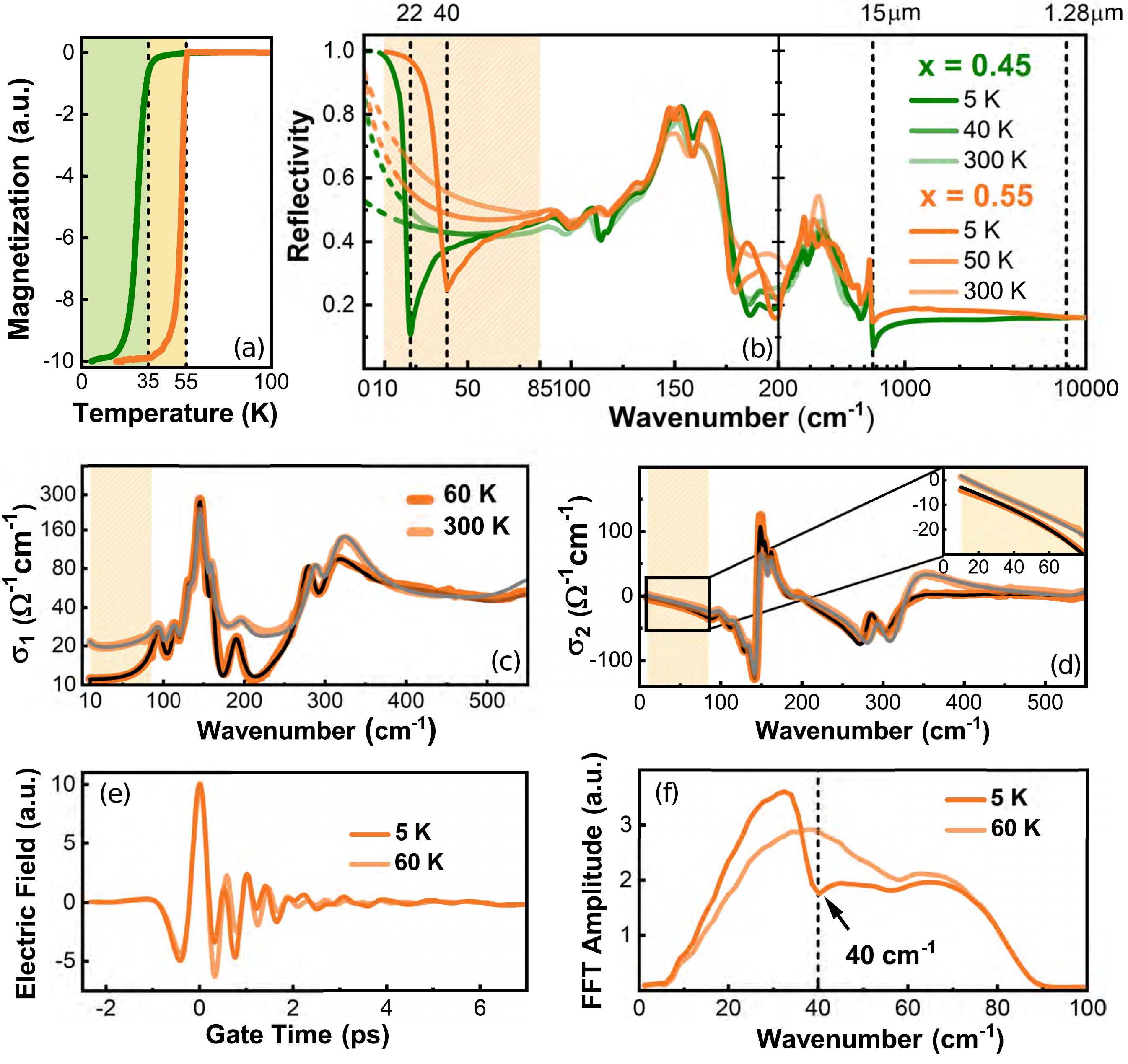}\\
  \caption{ Sample characterization and broadband reflectivity spectra along c-axis in equilibrium state. (a) Temperature dependent magnetization measurements. (b) Out-of-plane reflectance spectra of YBCO are governed by phonons in FIR region. In superconducting states, sharp Josephson plasmon edges show up in THz regime. Dashed lines are the low-frequency extrapolations used for Kramers-Kronig transformation. The grey shaded areas show the energy scale and spectral range of pump pluses. (c)-(d) The colored thick lines show the real and imaginary parts of conductivity in x=0.55 at 60 and 300K. The black and grey thin lines show the fitting results of Drude-Lorentz model. (e)-(f) Reflected THz electric field in time and frequency domain measured by a time-domain THz spectroscopy system, whose measurement range is plotted as a shaded area in (b)-(d). }\label{Fig:1}
\end{figure}

Aiming at addressing two critical issues on photo-induced effect in YBCO, that is, whether phonon pumping is essential in inducing the effect, and whether the induced edge-like shape in reflectance spectrum and 1/$\omega$-like dependence in imaginary part of conductivity could be unambiguously attributed to transient superconductivity, here we present a comprehensive NIR and MIR pump, c-axis terahertz probe measurement on underdoped YBCO superconductors. First, we investigate respectively the non-equilibrium states after NIR excitations below and above T$_c$, which suggest a strong depletion of superconducting condensate and significant increase of quasiparticles, respectively. Secondly, we show that similar transient responses are detected after tuning the pump to be resonant with the phonon near 650 \cm. We elaborate that the present study does not favor positive answers to those questions above.

Details of sample growth and spectral measurements are presented in Supplemental Material \cite{supp}. Figure \ref{Fig:1} (a) shows the temperature dependent magnetization data for the two crystals measured under the magnetic field of 50 Oe in a Quantum Design Physical Property Measurement System. They show sharp superconducting transition temperature at T$_c$ = 55 K (x = 0.55) and 35 K (x = 0.45), respectively. Figure \ref{Fig:1} (b)  presents the broadband reflectivity spectra R($\omega$) of the two samples along c-axis at three different temperatures: 5 K, just above T$_c$ and 300 K, which are measured by Fourier transform infrared (FTIR) spectrometers in combination with a time domain THz spectroscopy system (details are presented in Supplemental Material \cite{supp}).

Similar to the previous results on underdoped YBCO \cite{Homes1995,PhysRevB.55.6051}, the out-of-plane reflectivities are dominated by phonons in far-infrared (FIR) region. Among all those phonons, the one locates near 650 \cm, assigned as an A$_{2u}$ mode involving the bonds between apical oxgens and copper atoms \cite{homes1995optical}, has been drawing much attention for it may be related to the enhancement of superconductivity in YBCO \cite{mankowsky2014nonlinear}. Below 85 \cm, the reflectivity becomes smooth. An upward curvature suggests presence of free carrier contribution in normal state. The x=0.55 sample apparently has higher reflectivity and shows stronger upward curvature at very low frequency. With increasing temperature, one can also observe an increase of low energy reflectivity. On the other hand, in superconducting state, sharp JPEs develop with reflectivity minimums locating at 40 \cm for x=0.55 and 22.4 \cm for x=0.45, respectively.

Figure \ref{Fig:1} (c) and (d) show the real and imaginary parts of conductivity below 550 \cm, derived from Kramers-Kronig transformation of broadband reflectivity, for x=0.55 sample at 60 K and 300 K in normal state, respectively. An increase of low frequency spectral weight in $\sigma_1(\omega)$ from 60 K to 300 K is clearly observed, suggesting increased contribution from free carriers. The result is consistent with the change of reflectivity mentioned above. Concurrently, $\sigma_2(\omega)$ in low frequency shows an enhancement, as displayed in the inset of Fig. \ref{Fig:1} (d). The thin black and grey lines are fitting curves which we shall explain in the discussion part.

Figure \ref{Fig:1} (e) shows the waveforms from time-domain THz measurements on x=0.55 sample at two different temperatures. Figure \ref{Fig:1} (f) shows the Fourier transformation of the THz electric field. The oscillations observed in time-domain at 5 K lead to a strong dip feature at 40 \cm in frequency domain, which is in good agreement with the JPE dip observed in FTIR reflectance measurements.

In our pump-probe experiments, two selective pump wavelength, as indicated by grey shaded areas in Fig. \ref{Fig:1} (b), are used to interrogate the pump-induced change in THz regime. MIR pump pulses are tuned to 15 $\upmu$m (667 \cm) being resonant with the phonon near 650 \cm, and NIR pump pulses are set to 1.28 $\upmu$m (7810 \cm) whose energy scale is notably higher than all the phonons and collective modes in YBCO.

\begin{figure*}[htbp]
  \centering
\includegraphics[width=18cm, trim=0 0 0 0,clip]{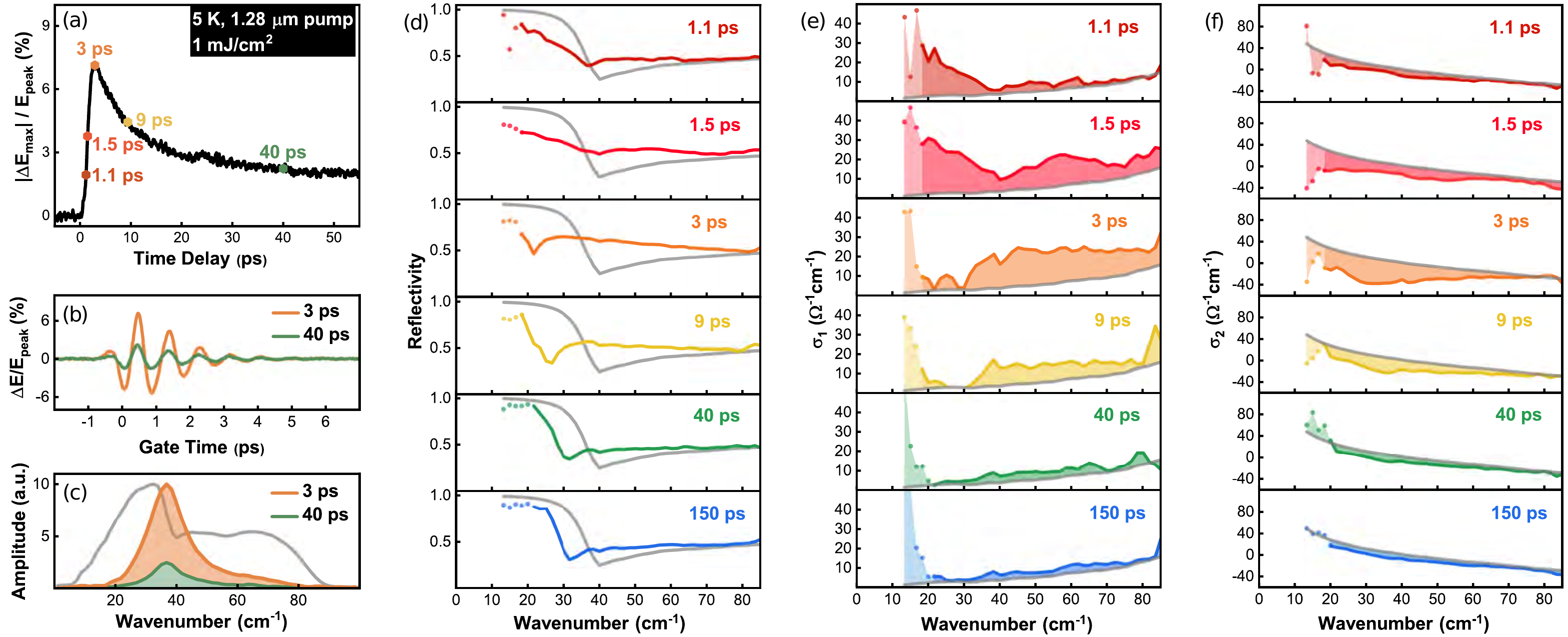}\\
  \caption{Pump-induced changes after excitation of 1.28 $\mu m$ at 5 K. (a)  the decay procedure of $|\Delta E_{max}|/E_{peak}$ after excited by a fluence of 1 mJ/cm$^2$ (peak electric filed of $\sim$3.9 MV/cm). $E_{peak}$ is the maximum value of ${E}_0(t)$ at static position. (b) the pump-induced relative change $\Delta{E}$(t, $\tau$)/$E_{peak}$ in time domain at 3 ps and 40 ps. (c) Fourier transformed spectrum of $\Delta{E}(t, \tau)$. Grey lines is the Fourier transformation of static reflected electric field divided by a coefficient. (d) transient reflectivity $R(\omega, \tau)$ (e) real part of conductivity $\sigma_{1}(\omega, \tau)$ (f) imaginary part of conductivity $\sigma_{2}(\omega, \tau)$. Some portions are plotted in color-fading dots for light-induced measurement signal there has almost vanishing spectral weight, which makes those calculated data points unreliable. Optical constants in static state are plotted in grey lines.  }\label{Fig:DeltaE}
\end{figure*}

We now present NIR pump-induced change in THz regime in superconducting state first. Because the two samples exhibit qualitatively the same behaviors, we show only the data collected on x=0.55 sample in the main text, while the whole set of data on x=0.45 sample are available in Supplemental Material \cite{supp}. Figure \ref{Fig:DeltaE} (a) displays the decay procedure of the maximum absolute value of pump-induced change, $|\Delta{E}_{max}|/E_{peak}$, after excitation at 1.28 $\mu m$ by a fluence of 1 mJ/cm$^2$ (peak electric filed of $\sim$3.9 MV/cm). We define the time zero at the position where $|\Delta{E}_{max}|/E_{peak}$ starts to change. Roughly a 7$\%$ maximum relative change is seen at 3 ps after excitation. The signal decays to $\sim$37$\%$ of the maximum signal within 40 ps before reaching a metastable state \cite{PhysRevB.98.020506,Cremin2019}. Then, the pump-induced relative change $\Delta {E}(t,\tau)/E_{peak}$ at time delay $\tau$ in superconducting state could be measured by fixing the THz gate line at the fixed delay time $\tau$. The methods for determining the phase of $\Delta{E}(t,\tau)$ relative to static $E(t)$ are presented in Supplemental Material \cite{supp}. Figure \ref{Fig:DeltaE} (b) shows $\Delta{E}(t,\tau)/E_{peak}$ at two representative delay time $\tau$ = 3 ps and 40 ps. The amplitudes of Fourier transformation of $\Delta{E}(t,\tau)$, i.e. $|\Delta{E}(\omega,\tau)|$, are displayed in Fig. \ref{Fig:DeltaE} (c). $|\Delta {E}(\omega,\tau)|$  behaves quite differently compared with the static reflected electric field of THz $|{E}_0(\omega)|$ (plotted in grey lines). The oscillations of pump-induced THz signal $\Delta {E}(t,\tau)/E_{peak}$ in time domain with an oscillation period of $\sim$0.85 ps give a pronounced peak near 40 \cm in frequency domain $|\Delta{E}(\omega,\tau)|$, suggesting that the pump-induced change occurs predominantly near the static JPE position.

A multilayer model is used to obtain the authentic pump-induced change of optical properties, which is disguised by the non-negligible mismatch of the penetration depth of pump and probe pulses. The detailed calculation method and rationality of the model are presented in Supplemental Material \cite{supp}. All the transient optical constants shown below are calculated with the multilayer model. It deserves to remark that, although the static electric field $|{E}_0(\omega)|$ has sufficiently high signal down to less than 10 \cm, the pump-induced signal of $|\Delta{E}(\omega,\tau)|$ in superconducting state changes predominantly near JPE and only has vanishing spectral weight below 18 \cm depending on time delays (see Fig. \ref{Fig:DeltaE} (c)), which makes the calculated optical constants dubious below 18 \cm. We distinguish those data from others with color-fading dots in Fig. \ref{Fig:DeltaE}.

Figure \ref{Fig:DeltaE} (d) shows the transient reflectivity at selective time delays. The NIR excitations dramatically change the low frequency c-axis response. Even before reaching the maximum pump-probe signal, e.g. at the time delay $\tau$=1.1 and 1.5 ps, reflectivity near JPE are strongly modified: the reflectivity values below static JPE are suppressed and that above JPE enhanced. The observations indicate a breakdown of Josephson tunneling along c-axis and development of quasiparticle excitations. At the maximum position of pump-induced signal, i.e. at 3 ps, a small edge re-appears at lower energy scale. Then the edge gets sharper at subsequent time delays and shifts towards static JPE at higher energy scale. This procedure reflects the recovery of Josephson tunneling after excitation.

The depression/recovery procedure of Josephson plasmon mode and quasiparticle excitations can be seen more clearly in real and imaginary parts of conductivity, i.e. $\sigma_1(\omega)$ and $\sigma_2(\omega)$ respectively. In static state, the values of $\sigma_1(\omega)$ along c-axis are rather low, which is shown as the grey lines in Fig. \ref{Fig:DeltaE} (e). Upon NIR excitations, the low frequency spectral weight develops in $\sigma_1(\omega)$, which reflects pump-induced quasiparticle excitations. The quasiparticle spectral weight increases sharply upon initial pumping, reaches maximum at the time delay close to the maximum signal, and then decreases at subsequent time delays. Meanwhile, $\sigma_2(\omega)$ deviates quickly from that in static state which roughly follows 1/$\omega$ dependence arising from the superconducting condensate, as shown in Fig. \ref{Fig:DeltaE} (f). The deviation becomes most prominent at the time delay near the maximum pumping signal, then gets smaller at further time delays. Those results strongly suggest that superconducting condensate is heavily disturbed or destroyed upon NIR pumping and gradually recovered with time delays.

\begin{figure*}[htbp]
  \centering
\includegraphics[width=17.5cm, trim=0 0 0 0,clip]{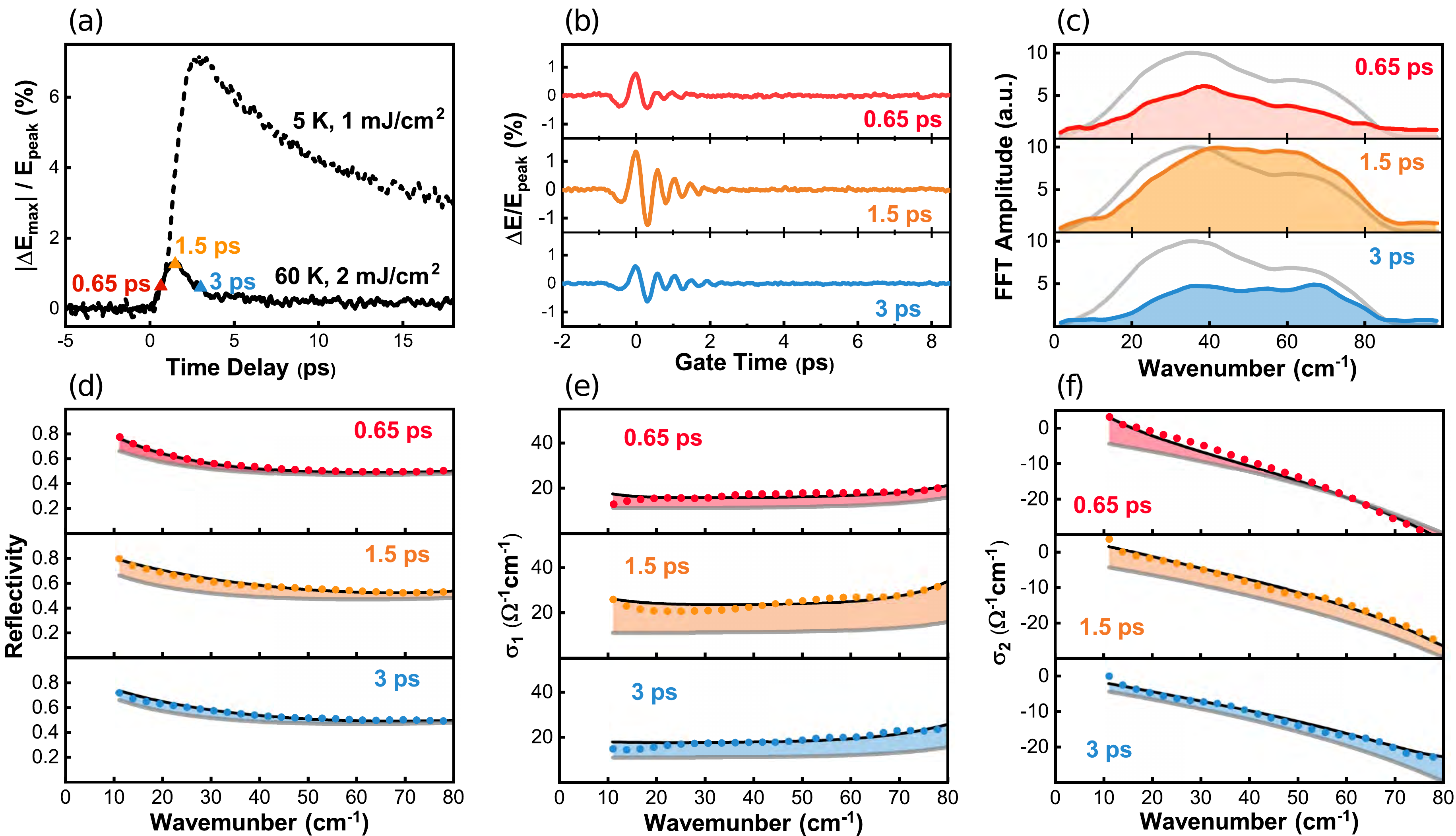}\\
  \caption{Pump-induced changes after excitation at 1.28 $\mu m$ by a fluence of 2 mJ/cm$^2$ (peak electric filed of $\sim$5.5 MV/cm) in normal state. (a)  the decay procedure of $|\Delta E_{max}|/E_{peak}$ at 60 K is plotted in a solid line. Three representative time delays are labeled in colored triangles. The decay procedure profile at 5 K is plotted in a dashed line for comparison. (b) the pump-induced relative change $\Delta{E}$(t, $\tau$)/$E_{peak}$ in time domain at three different time delays. (c) Fourier transformed spectrum of $\Delta{E}(t, \tau)$. Grey lines is Fourier transformation of static reflected electric field divided by a coefficient. (d)-(f) transient reflectivity $R(\omega, \tau)$, real part of conductivity $\sigma_{1}(\omega, \tau)$, imaginary part of conductivity $\sigma_{2}(\omega, \tau)$, respectively. The colored dots are experimental data and the black lines are fitting curves. Optical constants in static state are plotted in grey lines. }\label{Fig:3}
\end{figure*}

Figure \ref{Fig:3} summarizes the NIR pump-induced change after the excitation at 1.28 $\upmu$m by a fluence of 2 mJ/cm$^2$ (peak electric filed of $\sim$5.5 MV/cm) at the temperature of 60 K. Three representative time delays shown in Fig. \ref{Fig:3} (a) will be discussed. The middle panel of Fig. \ref{Fig:3} (b) shows the measured pump-induced change $\Delta {E}(t,\tau)$ at 1.5 ps, which represents the maximum photo-induced response in normal state. The upper and lower panels show $\Delta {E}(t,\tau)$ at 0.65 ps and 3 ps, which locate at half of maximum position of the rise and decay procedures, respectively. According to the raw experimental data in  Fig. \ref{Fig:3} (a) and  (b), there are essentially three differences in normal state compared with that in superconducting state. Firstly, the rise and decay time of $|\Delta E_{max}|/E_{peak}$ are relatively short in normal phase as shown in Fig. \ref{Fig:3} (a). Secondly, the pump-induced change of reflected electric field is significantly smaller in normal phase compared with that in superconducting state. Thirdly, the damped oscillations with an oscillation period of $\sim$0.85 ps observed in superconducting state at 5 K disappears completely in the pump-induced change $\Delta {E}(t,\tau)$ , and the time duration of $\Delta {E}(t,\tau)$ is within 3 ps in normal phase, as shown in Fig. \ref{Fig:3} (b). Figure \ref{Fig:3} (c) shows the amplitude of Fourier transformation of $\Delta {E}(t,\tau)$, which covers nearly all the THz regime generated in our experiments. $\Delta {E}(t,\tau)$ is multiplied by a Blackman window function before doing Fourier transformation to reduce the noise effect, which will not affect the calculated transient optical constants.

Figure \ref{Fig:3} (d) (e) and (f) present the transient optical constants calculated with the multilayer model at the three time delays. For all those three time delays after excitation, a slight enhancement of reflectivity can be observed. At 0.65 ps, the enhancement at lower frequency dominates more than that at higher. At the maximum pump-induced response position, 1.5 ps, transient reflectivity ascends collectively. Then the reflectivity turns to decay, as seen at 3 ps. Enhancement of low frequency reflectivity usually reflects enhanced contribution from free carriers, which is also supported by simultaneous rise of $\sigma_1(\omega)$ and $\sigma_2(\omega)$. It indicates that the NIR pump pulses turn to result in quasiparticle excitations. We shall address this issue further in the discussion part. The fitting curves depicted in Fig. \ref{Fig:3} (d) (e) and (f) will also be explained there.

We also perform MIR pump-THz probe experiments on YBCO in superconducting and normal states as shown in Fig. \ref{Fig:pump15mum}, in which the MIR pump pulses are tuned to 15 $\upmu$m and resonant with the apical-oxygen-related phonon mode. The MIR pump is at a fluence of 1 mJ/cm$^2$ and the peak electric field is $\sim$1.5 MV/cm. In both superconducting and normal state, the sample exhibits similar rise and decay procedures of $|\Delta E_{max}|/E_{peak}$ after the excitation by MIR (Fig. \ref{Fig:pump15mum} (a)) and NIR (Fig. \ref{Fig:3} (a)) pump pulses, though the decay of $|\Delta E_{max}|/E_{peak}$ after MIR excitation appears more significant. Figure \ref{Fig:pump15mum} (b) shows the waveform of pump-induced change of reflected THz electric field at maximum response position, $\tau$=3 ps, in superconducting state (T=5 K). Figure \ref{Fig:pump15mum} (c) (d) and (e) show the calculated transient optical constants, R($\omega$, $\tau$), $\sigma_1(\omega, \tau)$, $\sigma_2(\omega, \tau)$, using the above mentioned multilayer model. The transient responses by NIR excitations at same fluence and same time delay are also shown in those panels as dashed lines for comparison. Similar to NIR pump pulses, in superconducting state, MIR pulses also turn to remove JPE upon exciting and then drive YBCO into a state with JPE at lower energy scale together with some spectral weight increase arising from excited quasiparticles. Figure \ref{Fig:pump15mum} (f)-(i) show the transient responses at time delay $\tau$=1.5 ps by MIR excitations in normal state (60 K), together with those by NIR excitations. We find that MIR excitations also lead to an increase of reflectivity and enhancement in both real and imaginary parts of conductivity, implying quasiparticle excitations. Although the pump-induced changes after NIR excitations (dashed lines in Fig. \ref{Fig:pump15mum} (d) and (e)) are more significant than that after MIR pump, there is no essential difference between those two transient responses, as seen in Fig. \ref{Fig:pump15mum}.

\begin{figure}[htbp]
  \centering
\includegraphics[width=8.5cm, trim=0 0 0 0,clip]{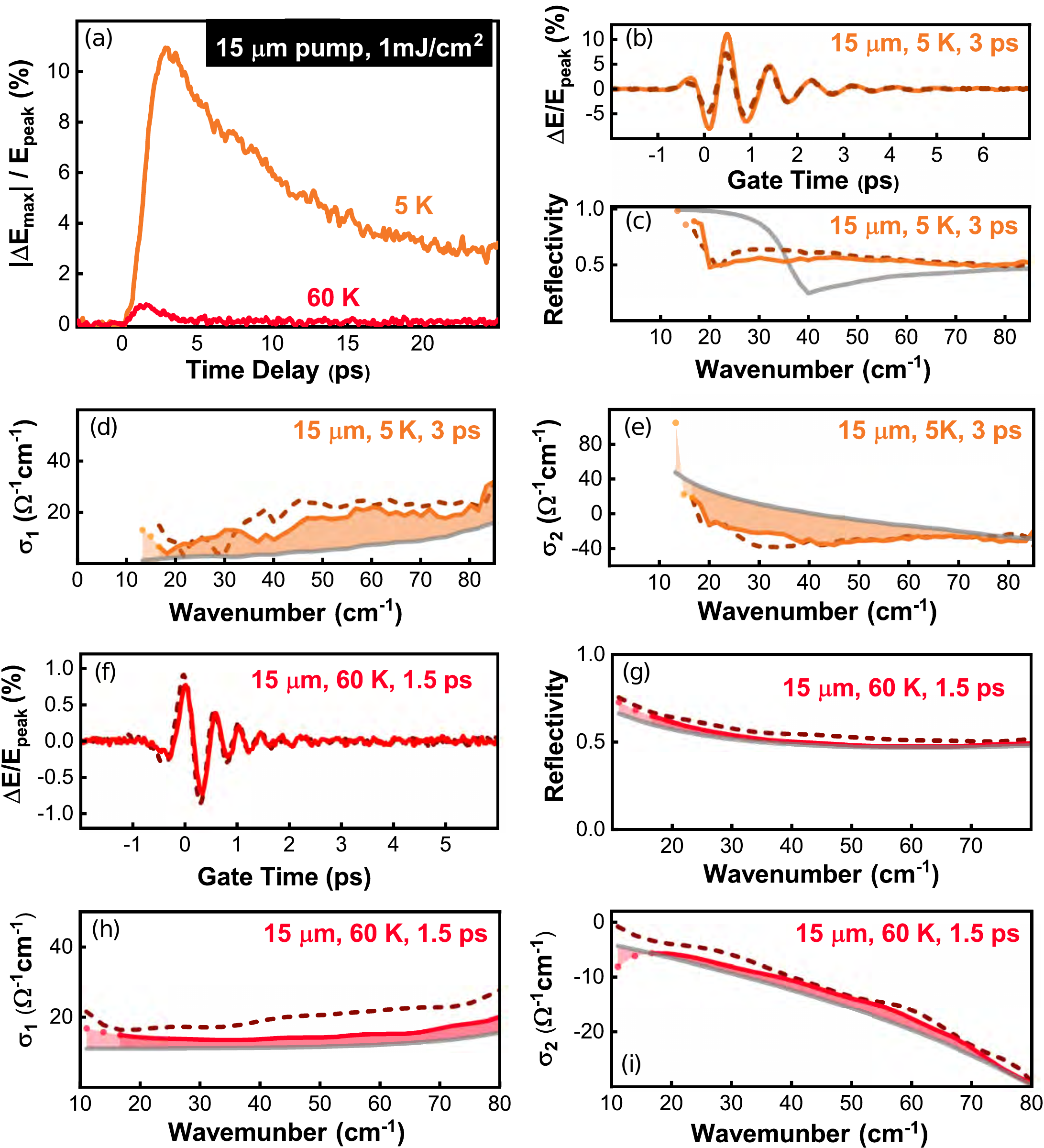}\\
  \caption{ Comparison between the pump-induced effects after the excitation of 15 $\upmu$m (solid lines) and 1.28 $\upmu$m (dashed lines) pulses at 1 mJ/cm$^2$. (a) the decay procedure of $|\Delta E_{max}|/E_{peak}$ at 5 K and 60 K. (b) the pump-induced change of THz electric field at 5 K in time domain.  (c)-(e) transient optical constants, R($\omega,\tau$), $\sigma_1(\omega,\tau)$, $\sigma_2(\omega,\tau)$ at 5 K. (f)  the pump-induced change of THz electric field at 60 K in time domain. (g)-(i) the transient optical constants at 60 K in normal state. Grey lines are the optical constants in the equilibrium state.}\label{Fig:pump15mum}

\end{figure}

Below we will make a brief comparison between the present experimental results with others reported in Ref. \citenum{hu2014optically, Kaiser2014, liu2019pump}. In superconduting state, the transient reflectivities at 1.1 ps and 1.5 ps are close to the reported results after excited by 15 $\upmu$m (MIR) at the fluence of 4 mJ/cm$^2$ and peak electric field of  $\sim$3 MV/cm (see Fig. 5j in Ref. \citenum{hu2014optically}). Both of those measurements show that the transient reflectivity is a gentle slope with a dip near the static JPE in superconducting state, which denotes a depletion of superconducting condensate and excitation of quasiparticles here. However, the observed feature was interpreted as the static JPE being fixed accompanied by a weak new edge appearing at higher energy scale \cite{hu2014optically, Kaiser2014}. In order to explain the edge appearing at higher energy, they assumed non-homogeneous excitations in the pumped regime even after considering the penetration depth mismatch. They claimed the superconductivity in a great portion (80\%) of the pumped volume stays unaffected with that in a small portion (20\%) being enhanced (see Fig. FS3.1 in supplementary information of Ref. \citenum{hu2014optically}). That explanation cannot be supported by our experiments. In fact, the spectral shape of transient reflectivity at any time delay presented here could not be reproduced by a combination of two different Josephson plasma edges with different volume fractions. Furthermore, the transient reflectivity at subsequent time depicts a re-emergence and recovery procedure of the static JPE, which indicates that suppression of the static JPE upon excitations is a more plausible explanation. Our result reveals that the Bruggeman's effective medium model is not applicable in the superconducting state. A recent report posted on arXiv by Liu et al. \cite{liu2019pump} confirms the spectral shape we observed in superconducting state when they use similar MIR pump fluence at 1.5 mJ/cm$^2$ (peak electric field at  $\sim$1.4 MV/cm), indicating a depletion of superconducting condensate. However, when they increase the MIR pump to higher fluence of 8 mJ/cm$^2$ and peak electric field of $\sim$3 MV/cm (see Fig. S8 in supplementary information of Ref. \citenum{liu2019pump}), they observe a transient increase in superfluid density in superconducting state. We remark that we could not make a comparison on MIR regime due to the limitation on MIR pump fluence in our experiments. Nevertheless, we performed measurement with NIR (1.28 $\upmu$m) excitation at higher fluence of 6 mJ/cm$^2$ (peak electric field of $\sim$9.5 MV/cm) in both superconducting and normal state, in which the peak electric field is much higher than other reports on YBCO. Rather than an enhancement/emergence of superconductivity, the results still indicate superconductivity depletion and quasiparticle excitations after strong NIR pump. Details can be found in Supplemental Material \cite{supp}.

The present work enables us to address two crucial issues with respect to the photo-induced transient superconductivity. The first one is whether or not phonon resonant pump is essential for the photo-induced effect? In the last few years, mode-selective optical control became particularly attractive because it is argued based on the MIR pump that the coherent excitation of certain phonon mode can suppress the charge order and simultaneously enhance superconductivity in cuprates. It has motivated many theoretical studies in this direction \cite{Subedi2014,PhysRevLett.114.137001,PhysRevB.91.184506,PhysRevLett.118.087002,Okamoto2017,Bittner2019,Klein}. Since we find qualitatively the same effect by NIR and MIR pump, phonon resonant pump could be essentially ruled out. Following our work, the recent results by Liu et al. also indicate that NIR pump pulses behave in a similar fashion and have significant photoexcitation effect on YBCO \cite{liu2019pump}. Moreover, we shall emphasize that the calculated transient optical constants using the multilayer model are quite sensitive to the estimated penetration depth of pump pulses. The pump-induced change turns to get more significant when the penetration depth of pump pulses gets more inconspicuous compared with that of probe one (find in Supplemental Material \cite{supp}). At the phonon mode position, the penetration depth gets to a minimum and changes quickly when tuning frequency deviated from the phonon, which could lead to a maximum photo-induced effects even when raw experimental data $\Delta E(t,\tau)/E_{peak}$ are nearly the same.

The second important issue is whether or not the observed phenomenon in normal state could be attributed to photo-induced transient superconductivity? We shall address this issue from the pump-induced change in both transient R($\omega$) and transient $\sigma_2(\omega)$. Our measurement indeed confirms the presence of photo-induced enhancement in transient $R(\omega, \tau)$ (or edge-like upturn in $\Delta R/R$ (see Supplemental Material \cite{supp} for data on x=0.45 sample)) by both NIR and MIR excitations. However, the enhancement or edge-like shape is rather weak in the calculated reflectance (Fig. \ref{Fig:3} (d)). The results are similar to the data reported previously \cite{Kaiser2014,hu2014optically,PhysRevB.94.224303}. However, the interpretation of the transient enhancement as the emergence of a new Josephson plasma edge relies on the analysis based on non-homogeneous excitations, even after taking the penetration depth mismatch of pump and probe pulses into account. A quantitative fit yields roughly 20$\%$ of volume fraction in the pumped regime being transformed into transient superconducting component and contributed to the edge like upturn\cite{hu2014optically}. We elaborated above that our measurement results are not compatible with the non-homogeneous excitations in superconducting state, then there is no reason that the non-homogeneous excitations could be present in normal state. On this basis, we can not attribute the observed enhancement or edge-like shape to the transient superconductivity which gives rise to a JPE in normal state.

We now discuss the transient enhancement in the imaginary part of conductivity. The superconductivity-related response manifests naturally in the complex conductivity. At zero temperature, the real part of conductivity of a superconductor is condensed into a delta function at zero frequency, and by Kramers-Kronig transformation, the imaginary part of conductivity goes as 1/$\omega$ dependence:
\begin{equation}
\sigma(\omega)=\sigma_1(\omega)+i\sigma_2(\omega)=\frac{\omega_{ps}^2}{8}\delta(0)+i\frac{\omega_{ps}^2}{4\pi\omega},
\label{chik}
\end{equation}
where $\omega_{ps}^2$ is the condensed plasma frequency and related to London penetration depth by $\omega_{ps}^2=1/\lambda^2_L$. According to this equation, the 1/$\omega$ dependence in $\sigma_2(\omega)$ could be taken as an indication of superconductivity, and the $4\pi\omega\cdot\sigma_2(\omega)$ at zero frequency limit could be used to estimate the superconducting condensate or London penetration depth. Recently, 1/$\omega$ dependence observed in transient $\sigma_2(\omega)$ was widely taken to be another evidence for the photo-induced transient superconductivity in literature \cite{Kaiser2014,hu2014optically,Nicoletti2014,Hunt2015,PhysRevB.94.224303,PhysRevLett.121.267003, liu2019pump}. We would like to remark that the observation of  1/$\omega$ dependence in $\sigma_2(\omega)$ within a very limited range of frequency is not a sufficient condition for the identification of superconductivity. When a compound changes from normal state to superconducting state, the low frequency spectral weight in $\sigma_1(\omega)$ must be reduced due to the condensate to the delta function at zero frequency. The missing spectral weight in $\sigma_1(\omega)$ can also be used to derive the superconducting condensate:
\begin{equation}
\omega_{ps}^2=8\int_0^\infty[\sigma_1(\omega,T>T_c)-\sigma_1(\omega,T\ll T_c)]d\omega.
\label{chik}
\end{equation}
Indeed, a loss of low-frequency spectral weight in $\sigma_1(\omega)$ has always been observed in cuprate superconductors across the  superconducting transitions in the equilibrium state, including measurements polarized along the c-axis \cite{Basov1999}. However, this was not the case for the above mentioned works on laser-induced response in normal state \cite{Kaiser2014,hu2014optically,Nicoletti2014,Hunt2015,PhysRevB.94.224303,PhysRevLett.121.267003, liu2019pump}. In those works, when a 1/$\omega$ dependence or an upward increase with decreasing frequency was identified in $\sigma_2(\omega)$, the low-frequency $\sigma_1(\omega)$ does not drop compared with the values in the static state. Instead, the low-frequency spectral weight of $\sigma_1(\omega)$  still shows an increase. The simultaneous enhancement in both transient $\sigma_2(\omega)$  and $\sigma_1(\omega)$ is in contradiction to superconducting condensate.

As a matter of fact, even for a simple Drude response, the imaginary part of conductivity $\sigma_2(\omega)$ could show approximately a 1/$\omega$ dependence if $\omega\gg1/\tau$ ($1/\tau$ can be considered as a scattering rate). $\sigma_2(\omega)$ starts to decrease only when $\omega$ becomes smaller than $1/\tau$. When the scattering rate is very small, it becomes hard to distinguish between a Drude response and a superconducting response solely from the frequency dependence of $\sigma_2(\omega)$ in the very limited THz measurement frequency range. As shown in Fig. \ref{Fig:1} (b), the reflectivities of the two samples in normal state are not completely flat but show a clear increase feature at low frequency limit, implying the presence of free carrier contribution to the c-axis optical response. The free carrier contribution increases with increasing doping levels as reflected in the two samples. Actually, in earlier study on the c-axis charge dynamics on La$_{1-x}$Sr$_x$CuO$_4$, the c-axis reflectance shows clearly upturn feature in THz regime upon increasing doping and temperature \cite{Uchida1996}. Similar changes were also observed for YBCO \cite{Homes1995,PhysRevB.55.6051}. It is well known that, for cuprate superconductors, the charge conduction along the c-axis is rather complex, which involves oxygen bonding with virtual Cu 4s orbitals \cite{Ioffe1998,Xiang1996,Hussey2003,Ioffe1998}. The c-axis hopping integral is strongly in-plane momentum dependent, being zero at nodal direction and maximum at the antinodal direction for the simple tetragonal cuprate system. Furthermore, the scattering rates of charge carriers contributing to the c-axis conductivity could be different at different region of the Fermi surface or Fermi arc. The presence of Cu-O chains in YBCO further complicates the conduction. Therefore, it is not expected that the quasiparticle contribution to the c-axis conductivity could be well explained by a simple Drude response. Nonetheless, the Drude response picture can still be used to explain the trend of low-frequency optical data contributed from the quasiparticles.

To analyze the low-frequency Drude-like increase feature in R($\omega$) and compare all optical constants in both equilibrium and non-equilibrium state more quantitatively, we try to fit R($\omega$), $\sigma_1(\omega)$ and $\sigma_2(\omega)$ simultaneously by using a Drude-Lorentz model
 \begin{equation}
\epsilon(\omega)=\epsilon_\infty-\sum_{i=1}{{\omega_{pi}^2}\over{\omega^2+i\omega/\tau_i}}+\sum_{j=1}{{S_j^2}{e^{i\theta_{j}}}\over{\omega_j^2-\omega^2-i\omega/\tau_j}}.
\label{chik}
\end{equation}
We find that the low frequency optical constants could be well reproduced simply by using two Drude components and a number of Lorents components for phonon peaks. As an example, Fig. \ref{Fig:1} (c) and (d) shows the Drude-Lorentz fit to the frequency dependence of $\sigma_1(\omega)$ and $\sigma_2(\omega)$ in equilibrium state for x=0.55 crystal at 60 K and 300 K in far-infrared region, respectively. In the inset of Fig. \ref{Fig:1} (d), we show the enlarged part of $\sigma_2(\omega)$ at low frequency, an enhancement of $\sigma_2(\omega)$ is seen simply at elevated temperature. We also find that the transient optical constants after excitations at 60 K in Fig.\ref{Fig:3} (d), (e) and (f), and even increasing the pump fluence to 6 mJ/cm$^2$ with the peak electric field being $\sim$9.5 MV/cm (shown in Supplemental Material \cite{supp}), could be approximatel reproduced simultaneously by the model with enhanced plasma frequencies and slightly reduced
scattering rates of Drude components. Although the fitting parameters (shown in Supplemental Material \cite{supp}) may only reflect the trend of free carrier evolution for the reasons mentioned above, the fitting clearly illustrates that the upward increase or a 1/$\omega$-like behavior in $\sigma_2(\omega)$ in the very narrow measured frequency range could not be uniquely explained by the transient superconductivity. Quasiparticle excitations offers an alternative way to explain the observed phenomenon, which could be similar with the results of some single-color pump-probe experiments \cite{PhysRevLett.82.4918,PhysRevLett.101.227001,PhysRevB.84.180507}.

We emphasize that, the optical constants, i.e. the enhancement in R($\omega$) or simultaneously in $\sigma_1(\omega)$ and $\sigma_2(\omega)$, after NIR or MIR excitations evolve in a way very similar to the increase of doping level or even increase of temperature, it is thus reasonable to attribute the effect to the contribution of photoexcited quasiparticles. We would like to point out that, in early studies, the quasiparticle scattering rate along the c-axis was believed to be very large, for the conductivity is very small. Since we observe clear Drude-like increase of upturn of R($\omega$) at very low frequency in the equilibrium state, it turns out that the normally assumed very large scattering rate may not be true. This issue should be further studied.

In summary, we observe a strong pump-induced spectral change in terahertz frequency, predominantly near the energy scale of Josephson plasmon edge below T$_c$. The edge is almost removed upon initial photo-excitations, indicating a collapse of superconducting condensate. After a short time delay we observe the reappearance of a JPE at frequency lower than the static JPE, whose feature becomes more pronounced and shifts slightly to higher energy scale with time delay envolving. Meanwhile, quasiparticle excitations develop and contribute to the spectral weight in the real part of conductivity. Above T$_c$, a much smaller pump-induced effect is detected. An enhancement or weak edge-like shape  develops in the reflectance spectrum, which also results in an increase of both real and imaginary parts of conductivity.  In addition, we find very different time scales for achieving the maximum pump-probe signal between T$<$T$_c$ and T$>$T$_c$. We elaborate that pump-induced effect above T$_c$ is unlikely to be explained by photo-induced transient superconductivity. In addition, there is no substantial difference between the near- and mid-infrared pump cases both in superconducting and normal state, which indicates that phonon resonant pump as a scenario for the photo-excitation effect can be excluded.

\begin{center}
\small{\textbf{ACKNOWLEDGMENTS}}
\end{center}
This work was supported by National Natural Science Foundation of China (No. 11888101), the National Key Research and Development Program of China (No. 2017YFA0302904, 2016YFA0300902, 2016YFA0300403).
\bibliographystyle{apsrev4-1}
\bibliography{Terahertz}

\begin{thebibliography}{41}%
\makeatletter
\providecommand \@ifxundefined [1]{%
 \@ifx{#1\undefined}
}%
\providecommand \@ifnum [1]{%
 \ifnum #1\expandafter \@firstoftwo
 \else \expandafter \@secondoftwo
 \fi
}%
\providecommand \@ifx [1]{%
 \ifx #1\expandafter \@firstoftwo
 \else \expandafter \@secondoftwo
 \fi
}%
\providecommand \natexlab [1]{#1}%
\providecommand \enquote  [1]{``#1''}%
\providecommand \bibnamefont  [1]{#1}%
\providecommand \bibfnamefont [1]{#1}%
\providecommand \citenamefont [1]{#1}%
\providecommand \href@noop [0]{\@secondoftwo}%
\providecommand \href [0]{\begingroup \@sanitize@url \@href}%
\providecommand \@href[1]{\@@startlink{#1}\@@href}%
\providecommand \@@href[1]{\endgroup#1\@@endlink}%
\providecommand \@sanitize@url [0]{\catcode `\\12\catcode `\$12\catcode
  `\&12\catcode `\#12\catcode `\^12\catcode `\_12\catcode `\%12\relax}%
\providecommand \@@startlink[1]{}%
\providecommand \@@endlink[0]{}%
\providecommand \url  [0]{\begingroup\@sanitize@url \@url }%
\providecommand \@url [1]{\endgroup\@href {#1}{\urlprefix }}%
\providecommand \urlprefix  [0]{URL }%
\providecommand \Eprint [0]{\href }%
\providecommand \doibase [0]{http://dx.doi.org/}%
\providecommand \selectlanguage [0]{\@gobble}%
\providecommand \bibinfo  [0]{\@secondoftwo}%
\providecommand \bibfield  [0]{\@secondoftwo}%
\providecommand \translation [1]{[#1]}%
\providecommand \BibitemOpen [0]{}%
\providecommand \bibitemStop [0]{}%
\providecommand \bibitemNoStop [0]{.\EOS\space}%
\providecommand \EOS [0]{\spacefactor3000\relax}%
\providecommand \BibitemShut  [1]{\csname bibitem#1\endcsname}%
\let\auto@bib@innerbib\@empty
\bibitem [{\citenamefont {Tamasaku}\ \emph {et~al.}(1992)\citenamefont
  {Tamasaku}, \citenamefont {Nakamura},\ and\ \citenamefont
  {Uchida}}]{PhysRevLett.69.1455}%
  \BibitemOpen
  \bibfield  {author} {\bibinfo {author} {\bibfnamefont {K.}~\bibnamefont
  {Tamasaku}}, \bibinfo {author} {\bibfnamefont {Y.}~\bibnamefont {Nakamura}},
  \ and\ \bibinfo {author} {\bibfnamefont {S.}~\bibnamefont {Uchida}},\ }\href
  {\doibase 10.1103/PhysRevLett.69.1455} {\bibfield  {journal} {\bibinfo
  {journal} {Phys. Rev. Lett.}\ }\textbf {\bibinfo {volume} {69}},\ \bibinfo
  {pages} {1455} (\bibinfo {year} {1992})}\BibitemShut {NoStop}%
\bibitem [{\citenamefont {Uchida}\ \emph {et~al.}(1996)\citenamefont {Uchida},
  \citenamefont {Tamasaku},\ and\ \citenamefont {Tajima}}]{Uchida1996}%
  \BibitemOpen
  \bibfield  {author} {\bibinfo {author} {\bibfnamefont {S.}~\bibnamefont
  {Uchida}}, \bibinfo {author} {\bibfnamefont {K.}~\bibnamefont {Tamasaku}}, \
  and\ \bibinfo {author} {\bibfnamefont {S.}~\bibnamefont {Tajima}},\ }\href
  {\doibase 10.1103/PhysRevB.53.14558} {\bibfield  {journal} {\bibinfo
  {journal} {Phys. Rev. B}\ }\textbf {\bibinfo {volume} {53}},\ \bibinfo
  {pages} {14558} (\bibinfo {year} {1996})}\BibitemShut {NoStop}%
\bibitem [{\citenamefont {Homes}\ \emph {et~al.}(1993)\citenamefont {Homes},
  \citenamefont {Timusk}, \citenamefont {Liang}, \citenamefont {Bonn},\ and\
  \citenamefont {Hardy}}]{PhysRevLett.71.1645}%
  \BibitemOpen
  \bibfield  {author} {\bibinfo {author} {\bibfnamefont {C.~C.}\ \bibnamefont
  {Homes}}, \bibinfo {author} {\bibfnamefont {T.}~\bibnamefont {Timusk}},
  \bibinfo {author} {\bibfnamefont {R.}~\bibnamefont {Liang}}, \bibinfo
  {author} {\bibfnamefont {D.~A.}\ \bibnamefont {Bonn}}, \ and\ \bibinfo
  {author} {\bibfnamefont {W.~N.}\ \bibnamefont {Hardy}},\ }\href {\doibase
  10.1103/PhysRevLett.71.1645} {\bibfield  {journal} {\bibinfo  {journal}
  {Phys. Rev. Lett.}\ }\textbf {\bibinfo {volume} {71}},\ \bibinfo {pages}
  {1645} (\bibinfo {year} {1993})}\BibitemShut {NoStop}%
\bibitem [{\citenamefont {Fausti}\ \emph {et~al.}(2011)\citenamefont {Fausti},
  \citenamefont {Tobey}, \citenamefont {Dean}, \citenamefont {Kaiser},
  \citenamefont {Dienst}, \citenamefont {Hoffmann}, \citenamefont {Pyon},
  \citenamefont {Takayama}, \citenamefont {Takagi},\ and\ \citenamefont
  {Cavalleri}}]{Fausti189}%
  \BibitemOpen
  \bibfield  {author} {\bibinfo {author} {\bibfnamefont {D.}~\bibnamefont
  {Fausti}}, \bibinfo {author} {\bibfnamefont {R.~I.}\ \bibnamefont {Tobey}},
  \bibinfo {author} {\bibfnamefont {N.}~\bibnamefont {Dean}}, \bibinfo {author}
  {\bibfnamefont {S.}~\bibnamefont {Kaiser}}, \bibinfo {author} {\bibfnamefont
  {A.}~\bibnamefont {Dienst}}, \bibinfo {author} {\bibfnamefont {M.~C.}\
  \bibnamefont {Hoffmann}}, \bibinfo {author} {\bibfnamefont {S.}~\bibnamefont
  {Pyon}}, \bibinfo {author} {\bibfnamefont {T.}~\bibnamefont {Takayama}},
  \bibinfo {author} {\bibfnamefont {H.}~\bibnamefont {Takagi}}, \ and\ \bibinfo
  {author} {\bibfnamefont {A.}~\bibnamefont {Cavalleri}},\ }\href {\doibase
  10.1126/science.1197294} {\bibfield  {journal} {\bibinfo  {journal}
  {Science}\ }\textbf {\bibinfo {volume} {331}},\ \bibinfo {pages} {189}
  (\bibinfo {year} {2011})}\BibitemShut {NoStop}%
\bibitem [{\citenamefont {Kaiser}\ \emph {et~al.}(2014)\citenamefont {Kaiser},
  \citenamefont {Hunt}, \citenamefont {Nicoletti}, \citenamefont {Hu},
  \citenamefont {Gierz}, \citenamefont {Liu}, \citenamefont {{Le Tacon}},
  \citenamefont {Loew}, \citenamefont {Haug}, \citenamefont {Keimer},\ and\
  \citenamefont {Cavalleri}}]{Kaiser2014}%
  \BibitemOpen
  \bibfield  {author} {\bibinfo {author} {\bibfnamefont {S.}~\bibnamefont
  {Kaiser}}, \bibinfo {author} {\bibfnamefont {C.~R.}\ \bibnamefont {Hunt}},
  \bibinfo {author} {\bibfnamefont {D.}~\bibnamefont {Nicoletti}}, \bibinfo
  {author} {\bibfnamefont {W.}~\bibnamefont {Hu}}, \bibinfo {author}
  {\bibfnamefont {I.}~\bibnamefont {Gierz}}, \bibinfo {author} {\bibfnamefont
  {H.~Y.}\ \bibnamefont {Liu}}, \bibinfo {author} {\bibfnamefont
  {M.}~\bibnamefont {{Le Tacon}}}, \bibinfo {author} {\bibfnamefont
  {T.}~\bibnamefont {Loew}}, \bibinfo {author} {\bibfnamefont {D.}~\bibnamefont
  {Haug}}, \bibinfo {author} {\bibfnamefont {B.}~\bibnamefont {Keimer}}, \ and\
  \bibinfo {author} {\bibfnamefont {A.}~\bibnamefont {Cavalleri}},\ }\href
  {\doibase 10.1103/PhysRevB.89.184516} {\bibfield  {journal} {\bibinfo
  {journal} {Phys. Rev. B}\ }\textbf {\bibinfo {volume} {89}},\ \bibinfo
  {pages} {184516} (\bibinfo {year} {2014})}\BibitemShut {NoStop}%
\bibitem [{\citenamefont {Hu}\ \emph {et~al.}(2014)\citenamefont {Hu},
  \citenamefont {Kaiser}, \citenamefont {Nicoletti}, \citenamefont {Hunt},
  \citenamefont {Gierz}, \citenamefont {Hoffmann}, \citenamefont {Le~Tacon},
  \citenamefont {Loew}, \citenamefont {Keimer},\ and\ \citenamefont
  {Cavalleri}}]{hu2014optically}%
  \BibitemOpen
  \bibfield  {author} {\bibinfo {author} {\bibfnamefont {W.}~\bibnamefont
  {Hu}}, \bibinfo {author} {\bibfnamefont {S.}~\bibnamefont {Kaiser}}, \bibinfo
  {author} {\bibfnamefont {D.}~\bibnamefont {Nicoletti}}, \bibinfo {author}
  {\bibfnamefont {C.~R.}\ \bibnamefont {Hunt}}, \bibinfo {author}
  {\bibfnamefont {I.}~\bibnamefont {Gierz}}, \bibinfo {author} {\bibfnamefont
  {M.~C.}\ \bibnamefont {Hoffmann}}, \bibinfo {author} {\bibfnamefont
  {M.}~\bibnamefont {Le~Tacon}}, \bibinfo {author} {\bibfnamefont
  {T.}~\bibnamefont {Loew}}, \bibinfo {author} {\bibfnamefont {B.}~\bibnamefont
  {Keimer}}, \ and\ \bibinfo {author} {\bibfnamefont {A.}~\bibnamefont
  {Cavalleri}},\ }\href {\doibase 10.1038/nmat3963} {\bibfield  {journal}
  {\bibinfo  {journal} {Nature Materials}\ }\textbf {\bibinfo {volume} {13}},\
  \bibinfo {pages} {705} (\bibinfo {year} {2014})}\BibitemShut {NoStop}%
\bibitem [{\citenamefont {Subedi}\ \emph {et~al.}(2014)\citenamefont {Subedi},
  \citenamefont {Cavalleri},\ and\ \citenamefont {Georges}}]{Subedi2014}%
  \BibitemOpen
  \bibfield  {author} {\bibinfo {author} {\bibfnamefont {A.}~\bibnamefont
  {Subedi}}, \bibinfo {author} {\bibfnamefont {A.}~\bibnamefont {Cavalleri}}, \
  and\ \bibinfo {author} {\bibfnamefont {A.}~\bibnamefont {Georges}},\ }\href
  {\doibase 10.1103/PhysRevB.89.220301} {\bibfield  {journal} {\bibinfo
  {journal} {Phys. Rev. B}\ }\textbf {\bibinfo {volume} {89}},\ \bibinfo
  {pages} {220301} (\bibinfo {year} {2014})}\BibitemShut {NoStop}%
\bibitem [{\citenamefont {Denny}\ \emph {et~al.}(2015)\citenamefont {Denny},
  \citenamefont {Clark}, \citenamefont {Laplace}, \citenamefont {Cavalleri},\
  and\ \citenamefont {Jaksch}}]{PhysRevLett.114.137001}%
  \BibitemOpen
  \bibfield  {author} {\bibinfo {author} {\bibfnamefont {S.~J.}\ \bibnamefont
  {Denny}}, \bibinfo {author} {\bibfnamefont {S.~R.}\ \bibnamefont {Clark}},
  \bibinfo {author} {\bibfnamefont {Y.}~\bibnamefont {Laplace}}, \bibinfo
  {author} {\bibfnamefont {A.}~\bibnamefont {Cavalleri}}, \ and\ \bibinfo
  {author} {\bibfnamefont {D.}~\bibnamefont {Jaksch}},\ }\href {\doibase
  10.1103/PhysRevLett.114.137001} {\bibfield  {journal} {\bibinfo  {journal}
  {Phys. Rev. Lett.}\ }\textbf {\bibinfo {volume} {114}},\ \bibinfo {pages}
  {137001} (\bibinfo {year} {2015})}\BibitemShut {NoStop}%
\bibitem [{\citenamefont {Raines}\ \emph {et~al.}(2015)\citenamefont {Raines},
  \citenamefont {Stanev},\ and\ \citenamefont {Galitski}}]{PhysRevB.91.184506}%
  \BibitemOpen
  \bibfield  {author} {\bibinfo {author} {\bibfnamefont {Z.~M.}\ \bibnamefont
  {Raines}}, \bibinfo {author} {\bibfnamefont {V.}~\bibnamefont {Stanev}}, \
  and\ \bibinfo {author} {\bibfnamefont {V.~M.}\ \bibnamefont {Galitski}},\
  }\href {\doibase 10.1103/PhysRevB.91.184506} {\bibfield  {journal} {\bibinfo
  {journal} {Phys. Rev. B}\ }\textbf {\bibinfo {volume} {91}},\ \bibinfo
  {pages} {184506} (\bibinfo {year} {2015})}\BibitemShut {NoStop}%
\bibitem [{\citenamefont {Sentef}\ \emph {et~al.}(2017)\citenamefont {Sentef},
  \citenamefont {Tokuno}, \citenamefont {Georges},\ and\ \citenamefont
  {Kollath}}]{PhysRevLett.118.087002}%
  \BibitemOpen
  \bibfield  {author} {\bibinfo {author} {\bibfnamefont {M.~A.}\ \bibnamefont
  {Sentef}}, \bibinfo {author} {\bibfnamefont {A.}~\bibnamefont {Tokuno}},
  \bibinfo {author} {\bibfnamefont {A.}~\bibnamefont {Georges}}, \ and\
  \bibinfo {author} {\bibfnamefont {C.}~\bibnamefont {Kollath}},\ }\href
  {\doibase 10.1103/PhysRevLett.118.087002} {\bibfield  {journal} {\bibinfo
  {journal} {Phys. Rev. Lett.}\ }\textbf {\bibinfo {volume} {118}},\ \bibinfo
  {pages} {087002} (\bibinfo {year} {2017})}\BibitemShut {NoStop}%
\bibitem [{\citenamefont {Okamoto}\ \emph {et~al.}(2017)\citenamefont
  {Okamoto}, \citenamefont {Hu}, \citenamefont {Cavalleri},\ and\ \citenamefont
  {Mathey}}]{Okamoto2017}%
  \BibitemOpen
  \bibfield  {author} {\bibinfo {author} {\bibfnamefont {J.}~\bibnamefont
  {Okamoto}}, \bibinfo {author} {\bibfnamefont {W.}~\bibnamefont {Hu}},
  \bibinfo {author} {\bibfnamefont {A.}~\bibnamefont {Cavalleri}}, \ and\
  \bibinfo {author} {\bibfnamefont {L.}~\bibnamefont {Mathey}},\ }\href
  {\doibase 10.1103/PhysRevB.96.144505} {\bibfield  {journal} {\bibinfo
  {journal} {Phys. Rev. B}\ }\textbf {\bibinfo {volume} {96}},\ \bibinfo
  {pages} {144505} (\bibinfo {year} {2017})}\BibitemShut {NoStop}%
\bibitem [{\citenamefont {Bittner}\ \emph {et~al.}(2019)\citenamefont
  {Bittner}, \citenamefont {Tohyama}, \citenamefont {Kaiser},\ and\
  \citenamefont {Manske}}]{Bittner2019}%
  \BibitemOpen
  \bibfield  {author} {\bibinfo {author} {\bibfnamefont {N.}~\bibnamefont
  {Bittner}}, \bibinfo {author} {\bibfnamefont {T.}~\bibnamefont {Tohyama}},
  \bibinfo {author} {\bibfnamefont {S.}~\bibnamefont {Kaiser}}, \ and\ \bibinfo
  {author} {\bibfnamefont {D.}~\bibnamefont {Manske}},\ }\href {\doibase
  10.7566/JPSJ.88.044704} {\bibfield  {journal} {\bibinfo  {journal} {Journal
  of the Physical Society of Japan}\ }\textbf {\bibinfo {volume} {88}},\
  \bibinfo {pages} {044704} (\bibinfo {year} {2019})}\BibitemShut {NoStop}%
\bibitem [{\citenamefont {Klein}\ \emph {et~al.}()\citenamefont {Klein},
  \citenamefont {Christensen},\ and\ \citenamefont {Fernandes}}]{Klein}%
  \BibitemOpen
  \bibfield  {author} {\bibinfo {author} {\bibfnamefont {A.}~\bibnamefont
  {Klein}}, \bibinfo {author} {\bibfnamefont {M.~H.}\ \bibnamefont
  {Christensen}}, \ and\ \bibinfo {author} {\bibfnamefont {R.~M.}\ \bibnamefont
  {Fernandes}},\ }\href@noop {} {\ }\Eprint
  {http://arxiv.org/abs/arXiv:1809.05600v1} {arXiv:arXiv:1809.05600v1}
  \BibitemShut {NoStop}%
\bibitem [{\citenamefont {Nicoletti}\ \emph {et~al.}(2014)\citenamefont
  {Nicoletti}, \citenamefont {Casandruc}, \citenamefont {Laplace},
  \citenamefont {Khanna}, \citenamefont {Hunt}, \citenamefont {Kaiser},
  \citenamefont {Dhesi}, \citenamefont {Gu}, \citenamefont {Hill},\ and\
  \citenamefont {Cavalleri}}]{Nicoletti2014}%
  \BibitemOpen
  \bibfield  {author} {\bibinfo {author} {\bibfnamefont {D.}~\bibnamefont
  {Nicoletti}}, \bibinfo {author} {\bibfnamefont {E.}~\bibnamefont
  {Casandruc}}, \bibinfo {author} {\bibfnamefont {Y.}~\bibnamefont {Laplace}},
  \bibinfo {author} {\bibfnamefont {V.}~\bibnamefont {Khanna}}, \bibinfo
  {author} {\bibfnamefont {C.~R.}\ \bibnamefont {Hunt}}, \bibinfo {author}
  {\bibfnamefont {S.}~\bibnamefont {Kaiser}}, \bibinfo {author} {\bibfnamefont
  {S.~S.}\ \bibnamefont {Dhesi}}, \bibinfo {author} {\bibfnamefont {G.~D.}\
  \bibnamefont {Gu}}, \bibinfo {author} {\bibfnamefont {J.~P.}\ \bibnamefont
  {Hill}}, \ and\ \bibinfo {author} {\bibfnamefont {A.}~\bibnamefont
  {Cavalleri}},\ }\href {\doibase 10.1103/PhysRevB.90.100503} {\bibfield
  {journal} {\bibinfo  {journal} {Phys. Rev. B}\ }\textbf {\bibinfo {volume}
  {90}},\ \bibinfo {pages} {100503(R)} (\bibinfo {year} {2014})}\BibitemShut
  {NoStop}%
\bibitem [{\citenamefont {Casandruc}\ \emph {et~al.}(2015)\citenamefont
  {Casandruc}, \citenamefont {Nicoletti}, \citenamefont {Rajasekaran},
  \citenamefont {Laplace}, \citenamefont {Khanna}, \citenamefont {Gu},
  \citenamefont {Hill},\ and\ \citenamefont {Cavalleri}}]{PhysRevB.91.174502}%
  \BibitemOpen
  \bibfield  {author} {\bibinfo {author} {\bibfnamefont {E.}~\bibnamefont
  {Casandruc}}, \bibinfo {author} {\bibfnamefont {D.}~\bibnamefont
  {Nicoletti}}, \bibinfo {author} {\bibfnamefont {S.}~\bibnamefont
  {Rajasekaran}}, \bibinfo {author} {\bibfnamefont {Y.}~\bibnamefont
  {Laplace}}, \bibinfo {author} {\bibfnamefont {V.}~\bibnamefont {Khanna}},
  \bibinfo {author} {\bibfnamefont {G.~D.}\ \bibnamefont {Gu}}, \bibinfo
  {author} {\bibfnamefont {J.~P.}\ \bibnamefont {Hill}}, \ and\ \bibinfo
  {author} {\bibfnamefont {A.}~\bibnamefont {Cavalleri}},\ }\href {\doibase
  10.1103/PhysRevB.91.174502} {\bibfield  {journal} {\bibinfo  {journal} {Phys.
  Rev. B}\ }\textbf {\bibinfo {volume} {91}},\ \bibinfo {pages} {174502}
  (\bibinfo {year} {2015})}\BibitemShut {NoStop}%
\bibitem [{\citenamefont {Zhang}\ \emph
  {et~al.}(2018{\natexlab{a}})\citenamefont {Zhang}, \citenamefont {Wang},
  \citenamefont {Shi}, \citenamefont {Lin}, \citenamefont {Zhang},
  \citenamefont {Gu}, \citenamefont {Dong},\ and\ \citenamefont
  {Wang}}]{PhysRevB.98.020506}%
  \BibitemOpen
  \bibfield  {author} {\bibinfo {author} {\bibfnamefont {S.~J.}\ \bibnamefont
  {Zhang}}, \bibinfo {author} {\bibfnamefont {Z.~X.}\ \bibnamefont {Wang}},
  \bibinfo {author} {\bibfnamefont {L.~Y.}\ \bibnamefont {Shi}}, \bibinfo
  {author} {\bibfnamefont {T.}~\bibnamefont {Lin}}, \bibinfo {author}
  {\bibfnamefont {M.~Y.}\ \bibnamefont {Zhang}}, \bibinfo {author}
  {\bibfnamefont {G.~D.}\ \bibnamefont {Gu}}, \bibinfo {author} {\bibfnamefont
  {T.}~\bibnamefont {Dong}}, \ and\ \bibinfo {author} {\bibfnamefont {N.~L.}\
  \bibnamefont {Wang}},\ }\href {\doibase 10.1103/PhysRevB.98.020506}
  {\bibfield  {journal} {\bibinfo  {journal} {Phys. Rev. B}\ }\textbf {\bibinfo
  {volume} {98}},\ \bibinfo {pages} {020506(R)} (\bibinfo {year}
  {2018}{\natexlab{a}})}\BibitemShut {NoStop}%
\bibitem [{\citenamefont {Zhang}\ \emph
  {et~al.}(2018{\natexlab{b}})\citenamefont {Zhang}, \citenamefont {Wang},
  \citenamefont {Wu}, \citenamefont {Liu}, \citenamefont {Shi}, \citenamefont
  {Lin}, \citenamefont {Li}, \citenamefont {Dai}, \citenamefont {Dong},\ and\
  \citenamefont {Wang}}]{PhysRevB.98.224507}%
  \BibitemOpen
  \bibfield  {author} {\bibinfo {author} {\bibfnamefont {S.~J.}\ \bibnamefont
  {Zhang}}, \bibinfo {author} {\bibfnamefont {Z.~X.}\ \bibnamefont {Wang}},
  \bibinfo {author} {\bibfnamefont {D.}~\bibnamefont {Wu}}, \bibinfo {author}
  {\bibfnamefont {Q.~M.}\ \bibnamefont {Liu}}, \bibinfo {author} {\bibfnamefont
  {L.~Y.}\ \bibnamefont {Shi}}, \bibinfo {author} {\bibfnamefont
  {T.}~\bibnamefont {Lin}}, \bibinfo {author} {\bibfnamefont {S.~L.}\
  \bibnamefont {Li}}, \bibinfo {author} {\bibfnamefont {P.~C.}\ \bibnamefont
  {Dai}}, \bibinfo {author} {\bibfnamefont {T.}~\bibnamefont {Dong}}, \ and\
  \bibinfo {author} {\bibfnamefont {N.~L.}\ \bibnamefont {Wang}},\ }\href
  {\doibase 10.1103/PhysRevB.98.224507} {\bibfield  {journal} {\bibinfo
  {journal} {Phys. Rev. B}\ }\textbf {\bibinfo {volume} {98}},\ \bibinfo
  {pages} {224507} (\bibinfo {year} {2018}{\natexlab{b}})}\BibitemShut
  {NoStop}%
\bibitem [{\citenamefont {Nicoletti}\ \emph {et~al.}(2018)\citenamefont
  {Nicoletti}, \citenamefont {Fu}, \citenamefont {Mehio}, \citenamefont
  {Moore}, \citenamefont {Disa}, \citenamefont {Gu},\ and\ \citenamefont
  {Cavalleri}}]{PhysRevLett.121.267003}%
  \BibitemOpen
  \bibfield  {author} {\bibinfo {author} {\bibfnamefont {D.}~\bibnamefont
  {Nicoletti}}, \bibinfo {author} {\bibfnamefont {D.}~\bibnamefont {Fu}},
  \bibinfo {author} {\bibfnamefont {O.}~\bibnamefont {Mehio}}, \bibinfo
  {author} {\bibfnamefont {S.}~\bibnamefont {Moore}}, \bibinfo {author}
  {\bibfnamefont {A.~S.}\ \bibnamefont {Disa}}, \bibinfo {author}
  {\bibfnamefont {G.~D.}\ \bibnamefont {Gu}}, \ and\ \bibinfo {author}
  {\bibfnamefont {A.}~\bibnamefont {Cavalleri}},\ }\href {\doibase
  10.1103/PhysRevLett.121.267003} {\bibfield  {journal} {\bibinfo  {journal}
  {Phys. Rev. Lett.}\ }\textbf {\bibinfo {volume} {121}},\ \bibinfo {pages}
  {267003} (\bibinfo {year} {2018})}\BibitemShut {NoStop}%
\bibitem [{\citenamefont {Cremin}\ \emph {et~al.}(2019)\citenamefont {Cremin},
  \citenamefont {Zhang}, \citenamefont {Homes}, \citenamefont {Gu},
  \citenamefont {Sun}, \citenamefont {Fogler}, \citenamefont {Millis},
  \citenamefont {Basov},\ and\ \citenamefont {Averitt}}]{Cremin2019}%
  \BibitemOpen
  \bibfield  {author} {\bibinfo {author} {\bibfnamefont {K.~A.}\ \bibnamefont
  {Cremin}}, \bibinfo {author} {\bibfnamefont {J.}~\bibnamefont {Zhang}},
  \bibinfo {author} {\bibfnamefont {C.~C.}\ \bibnamefont {Homes}}, \bibinfo
  {author} {\bibfnamefont {G.~D.}\ \bibnamefont {Gu}}, \bibinfo {author}
  {\bibfnamefont {Z.}~\bibnamefont {Sun}}, \bibinfo {author} {\bibfnamefont
  {M.~M.}\ \bibnamefont {Fogler}}, \bibinfo {author} {\bibfnamefont {A.~J.}\
  \bibnamefont {Millis}}, \bibinfo {author} {\bibfnamefont {D.~N.}\
  \bibnamefont {Basov}}, \ and\ \bibinfo {author} {\bibfnamefont {R.~D.}\
  \bibnamefont {Averitt}},\ }\href {\doibase 10.1073/pnas.1908368116}
  {\bibfield  {journal} {\bibinfo  {journal} {Proceedings of the National
  Academy of Sciences}\ }\textbf {\bibinfo {volume} {116}},\ \bibinfo {pages}
  {19875} (\bibinfo {year} {2019})},\ \Eprint
  {http://arxiv.org/abs/https://www.pnas.org/content/116/40/19875.full.pdf}
  {https://www.pnas.org/content/116/40/19875.full.pdf} \BibitemShut {NoStop}%
\bibitem [{\citenamefont {Niwa}\ \emph {et~al.}(2019)\citenamefont {Niwa},
  \citenamefont {Yoshikawa}, \citenamefont {Tomari}, \citenamefont {Matsunaga},
  \citenamefont {Song}, \citenamefont {Eisaki},\ and\ \citenamefont
  {Shimano}}]{PhysRevB.100.104507}%
  \BibitemOpen
  \bibfield  {author} {\bibinfo {author} {\bibfnamefont {H.}~\bibnamefont
  {Niwa}}, \bibinfo {author} {\bibfnamefont {N.}~\bibnamefont {Yoshikawa}},
  \bibinfo {author} {\bibfnamefont {K.}~\bibnamefont {Tomari}}, \bibinfo
  {author} {\bibfnamefont {R.}~\bibnamefont {Matsunaga}}, \bibinfo {author}
  {\bibfnamefont {D.}~\bibnamefont {Song}}, \bibinfo {author} {\bibfnamefont
  {H.}~\bibnamefont {Eisaki}}, \ and\ \bibinfo {author} {\bibfnamefont
  {R.}~\bibnamefont {Shimano}},\ }\href {\doibase 10.1103/PhysRevB.100.104507}
  {\bibfield  {journal} {\bibinfo  {journal} {Phys. Rev. B}\ }\textbf {\bibinfo
  {volume} {100}},\ \bibinfo {pages} {104507} (\bibinfo {year}
  {2019})}\BibitemShut {NoStop}%
\bibitem [{\citenamefont {Homes}\ \emph
  {et~al.}(1995{\natexlab{a}})\citenamefont {Homes}, \citenamefont {Timusk},
  \citenamefont {Bonn}, \citenamefont {Liang},\ and\ \citenamefont
  {Hardy}}]{Homes1995}%
  \BibitemOpen
  \bibfield  {author} {\bibinfo {author} {\bibfnamefont {C.~C.}\ \bibnamefont
  {Homes}}, \bibinfo {author} {\bibfnamefont {T.}~\bibnamefont {Timusk}},
  \bibinfo {author} {\bibfnamefont {D.~A.}\ \bibnamefont {Bonn}}, \bibinfo
  {author} {\bibfnamefont {R.}~\bibnamefont {Liang}}, \ and\ \bibinfo {author}
  {\bibfnamefont {W.~N.}\ \bibnamefont {Hardy}},\ }\href {\doibase
  10.1139/p95-099} {\bibfield  {journal} {\bibinfo  {journal} {Canadian Journal
  of Physics}\ }\textbf {\bibinfo {volume} {73}},\ \bibinfo {pages} {663}
  (\bibinfo {year} {1995}{\natexlab{a}})},\ \Eprint
  {http://arxiv.org/abs/https://doi.org/10.1139/p95-099}
  {https://doi.org/10.1139/p95-099} \BibitemShut {NoStop}%
\bibitem [{\citenamefont {Tajima}\ \emph {et~al.}(1997)\citenamefont {Tajima},
  \citenamefont {Sch\"utzmann}, \citenamefont {Miyamoto}, \citenamefont
  {Terasaki}, \citenamefont {Sato},\ and\ \citenamefont
  {Hauff}}]{PhysRevB.55.6051}%
  \BibitemOpen
  \bibfield  {author} {\bibinfo {author} {\bibfnamefont {S.}~\bibnamefont
  {Tajima}}, \bibinfo {author} {\bibfnamefont {J.}~\bibnamefont
  {Sch\"utzmann}}, \bibinfo {author} {\bibfnamefont {S.}~\bibnamefont
  {Miyamoto}}, \bibinfo {author} {\bibfnamefont {I.}~\bibnamefont {Terasaki}},
  \bibinfo {author} {\bibfnamefont {Y.}~\bibnamefont {Sato}}, \ and\ \bibinfo
  {author} {\bibfnamefont {R.}~\bibnamefont {Hauff}},\ }\href {\doibase
  10.1103/PhysRevB.55.6051} {\bibfield  {journal} {\bibinfo  {journal} {Phys.
  Rev. B}\ }\textbf {\bibinfo {volume} {55}},\ \bibinfo {pages} {6051}
  (\bibinfo {year} {1997})}\BibitemShut {NoStop}%
\bibitem [{\citenamefont {Homes}\ \emph
  {et~al.}(1995{\natexlab{b}})\citenamefont {Homes}, \citenamefont {Timusk},
  \citenamefont {Bonn}, \citenamefont {Liang},\ and\ \citenamefont
  {Hardy}}]{homes1995optical}%
  \BibitemOpen
  \bibfield  {author} {\bibinfo {author} {\bibfnamefont {C.}~\bibnamefont
  {Homes}}, \bibinfo {author} {\bibfnamefont {T.}~\bibnamefont {Timusk}},
  \bibinfo {author} {\bibfnamefont {D.}~\bibnamefont {Bonn}}, \bibinfo {author}
  {\bibfnamefont {R.}~\bibnamefont {Liang}}, \ and\ \bibinfo {author}
  {\bibfnamefont {W.}~\bibnamefont {Hardy}},\ }\href@noop {} {\bibfield
  {journal} {\bibinfo  {journal} {Canadian journal of physics}\ }\textbf
  {\bibinfo {volume} {73}},\ \bibinfo {pages} {663} (\bibinfo {year}
  {1995}{\natexlab{b}})}\BibitemShut {NoStop}%
\bibitem [{\citenamefont {Mankowsky}\ \emph {et~al.}(2014)\citenamefont
  {Mankowsky}, \citenamefont {Subedi}, \citenamefont {F{\"o}rst}, \citenamefont
  {Mariager}, \citenamefont {Chollet}, \citenamefont {Lemke}, \citenamefont
  {Robinson}, \citenamefont {Glownia}, \citenamefont {Minitti}, \citenamefont
  {Frano} \emph {et~al.}}]{mankowsky2014nonlinear}%
  \BibitemOpen
  \bibfield  {author} {\bibinfo {author} {\bibfnamefont {R.}~\bibnamefont
  {Mankowsky}}, \bibinfo {author} {\bibfnamefont {A.}~\bibnamefont {Subedi}},
  \bibinfo {author} {\bibfnamefont {M.}~\bibnamefont {F{\"o}rst}}, \bibinfo
  {author} {\bibfnamefont {S.~O.}\ \bibnamefont {Mariager}}, \bibinfo {author}
  {\bibfnamefont {M.}~\bibnamefont {Chollet}}, \bibinfo {author} {\bibfnamefont
  {H.}~\bibnamefont {Lemke}}, \bibinfo {author} {\bibfnamefont {J.~S.}\
  \bibnamefont {Robinson}}, \bibinfo {author} {\bibfnamefont {J.~M.}\
  \bibnamefont {Glownia}}, \bibinfo {author} {\bibfnamefont {M.~P.}\
  \bibnamefont {Minitti}}, \bibinfo {author} {\bibfnamefont {A.}~\bibnamefont
  {Frano}},  \emph {et~al.},\ }\href@noop {} {\bibfield  {journal} {\bibinfo
  {journal} {Nature}\ }\textbf {\bibinfo {volume} {516}},\ \bibinfo {pages}
  {71} (\bibinfo {year} {2014})}\BibitemShut {NoStop}%
\bibitem [{\citenamefont {Liu}\ \emph {et~al.}(2019)\citenamefont {Liu},
  \citenamefont {Forst}, \citenamefont {Fechner}, \citenamefont {Nicoletti},
  \citenamefont {Porras}, \citenamefont {Keimer},\ and\ \citenamefont
  {Cavalleri}}]{liu2019pump}%
  \BibitemOpen
  \bibfield  {author} {\bibinfo {author} {\bibfnamefont {B.}~\bibnamefont
  {Liu}}, \bibinfo {author} {\bibfnamefont {M.}~\bibnamefont {Forst}}, \bibinfo
  {author} {\bibfnamefont {M.}~\bibnamefont {Fechner}}, \bibinfo {author}
  {\bibfnamefont {D.}~\bibnamefont {Nicoletti}}, \bibinfo {author}
  {\bibfnamefont {J.}~\bibnamefont {Porras}}, \bibinfo {author} {\bibfnamefont
  {B.}~\bibnamefont {Keimer}}, \ and\ \bibinfo {author} {\bibfnamefont
  {A.}~\bibnamefont {Cavalleri}},\ }\href@noop {} {\enquote {\bibinfo {title}
  {Two pump frequency resonances for light-induced superconductivity in
  $\mathrm{Y}\mathrm{B}{\mathrm{a}}_{2}\mathrm{Cu}_{3}\mathrm{O}_{6.5}$},}\ }
  (\bibinfo {year} {2019}),\ \Eprint {http://arxiv.org/abs/1905.08356}
  {arXiv:1905.08356 [cond-mat.supr-con]} \BibitemShut {NoStop}%
\bibitem [{\citenamefont {Hunt}\ \emph {et~al.}(2016)\citenamefont {Hunt},
  \citenamefont {Nicoletti}, \citenamefont {Kaiser}, \citenamefont {Pr\"opper},
  \citenamefont {Loew}, \citenamefont {Porras}, \citenamefont {Keimer},\ and\
  \citenamefont {Cavalleri}}]{PhysRevB.94.224303}%
  \BibitemOpen
  \bibfield  {author} {\bibinfo {author} {\bibfnamefont {C.~R.}\ \bibnamefont
  {Hunt}}, \bibinfo {author} {\bibfnamefont {D.}~\bibnamefont {Nicoletti}},
  \bibinfo {author} {\bibfnamefont {S.}~\bibnamefont {Kaiser}}, \bibinfo
  {author} {\bibfnamefont {D.}~\bibnamefont {Pr\"opper}}, \bibinfo {author}
  {\bibfnamefont {T.}~\bibnamefont {Loew}}, \bibinfo {author} {\bibfnamefont
  {J.}~\bibnamefont {Porras}}, \bibinfo {author} {\bibfnamefont
  {B.}~\bibnamefont {Keimer}}, \ and\ \bibinfo {author} {\bibfnamefont
  {A.}~\bibnamefont {Cavalleri}},\ }\href {\doibase 10.1103/PhysRevB.94.224303}
  {\bibfield  {journal} {\bibinfo  {journal} {Phys. Rev. B}\ }\textbf {\bibinfo
  {volume} {94}},\ \bibinfo {pages} {224303} (\bibinfo {year}
  {2016})}\BibitemShut {NoStop}%
\bibitem [{\citenamefont {Hunt}\ \emph {et~al.}(2015)\citenamefont {Hunt},
  \citenamefont {Nicoletti}, \citenamefont {Kaiser}, \citenamefont {Takayama},
  \citenamefont {Takagi},\ and\ \citenamefont {Cavalleri}}]{Hunt2015}%
  \BibitemOpen
  \bibfield  {author} {\bibinfo {author} {\bibfnamefont {C.~R.}\ \bibnamefont
  {Hunt}}, \bibinfo {author} {\bibfnamefont {D.}~\bibnamefont {Nicoletti}},
  \bibinfo {author} {\bibfnamefont {S.}~\bibnamefont {Kaiser}}, \bibinfo
  {author} {\bibfnamefont {T.}~\bibnamefont {Takayama}}, \bibinfo {author}
  {\bibfnamefont {H.}~\bibnamefont {Takagi}}, \ and\ \bibinfo {author}
  {\bibfnamefont {A.}~\bibnamefont {Cavalleri}},\ }\href {\doibase
  10.1103/PhysRevB.91.020505} {\bibfield  {journal} {\bibinfo  {journal} {Phys.
  Rev. B}\ }\textbf {\bibinfo {volume} {91}},\ \bibinfo {pages} {020505}
  (\bibinfo {year} {2015})}\BibitemShut {NoStop}%
\bibitem [{\citenamefont {Basov}\ \emph {et~al.}(1999)\citenamefont {Basov},
  \citenamefont {Woods}, \citenamefont {Katz}, \citenamefont {Singley},
  \citenamefont {Dynes}, \citenamefont {Xu}, \citenamefont {Hinks},
  \citenamefont {Homes},\ and\ \citenamefont {Strongin}}]{Basov1999}%
  \BibitemOpen
  \bibfield  {author} {\bibinfo {author} {\bibfnamefont {D.~N.}\ \bibnamefont
  {Basov}}, \bibinfo {author} {\bibfnamefont {S.~I.}\ \bibnamefont {Woods}},
  \bibinfo {author} {\bibfnamefont {A.~S.}\ \bibnamefont {Katz}}, \bibinfo
  {author} {\bibfnamefont {E.~J.}\ \bibnamefont {Singley}}, \bibinfo {author}
  {\bibfnamefont {R.~C.}\ \bibnamefont {Dynes}}, \bibinfo {author}
  {\bibfnamefont {M.}~\bibnamefont {Xu}}, \bibinfo {author} {\bibfnamefont
  {D.~G.}\ \bibnamefont {Hinks}}, \bibinfo {author} {\bibfnamefont {C.~C.}\
  \bibnamefont {Homes}}, \ and\ \bibinfo {author} {\bibfnamefont
  {M.}~\bibnamefont {Strongin}},\ }\href {\doibase 10.1126/science.283.5398.49}
  {\bibfield  {journal} {\bibinfo  {journal} {Science}\ }\textbf {\bibinfo
  {volume} {283}},\ \bibinfo {pages} {49} (\bibinfo {year} {1999})}\BibitemShut
  {NoStop}%
\bibitem [{\citenamefont {Ioffe}\ and\ \citenamefont
  {Millis}(1998)}]{Ioffe1998}%
  \BibitemOpen
  \bibfield  {author} {\bibinfo {author} {\bibfnamefont {L.~B.}\ \bibnamefont
  {Ioffe}}\ and\ \bibinfo {author} {\bibfnamefont {A.~J.}\ \bibnamefont
  {Millis}},\ }\href {\doibase 10.1103/PhysRevB.58.11631} {\bibfield  {journal}
  {\bibinfo  {journal} {Phys. Rev. B}\ }\textbf {\bibinfo {volume} {58}},\
  \bibinfo {pages} {11631} (\bibinfo {year} {1998})}\BibitemShut {NoStop}%
\bibitem [{\citenamefont {Xiang}\ and\ \citenamefont
  {Wheatley}(1996)}]{Xiang1996}%
  \BibitemOpen
  \bibfield  {author} {\bibinfo {author} {\bibfnamefont {T.}~\bibnamefont
  {Xiang}}\ and\ \bibinfo {author} {\bibfnamefont {J.~M.}\ \bibnamefont
  {Wheatley}},\ }\href {\doibase 10.1103/PhysRevLett.77.4632} {\bibfield
  {journal} {\bibinfo  {journal} {Phys. Rev. Lett.}\ }\textbf {\bibinfo
  {volume} {77}},\ \bibinfo {pages} {4632} (\bibinfo {year}
  {1996})}\BibitemShut {NoStop}%
\bibitem [{\citenamefont {Hussey}\ \emph {et~al.}(2003)\citenamefont {Hussey},
  \citenamefont {Abdel-Jawad}, \citenamefont {Carrington}, \citenamefont
  {Mackenzie},\ and\ \citenamefont {Balicas}}]{Hussey2003}%
  \BibitemOpen
  \bibfield  {author} {\bibinfo {author} {\bibfnamefont {N.~E.}\ \bibnamefont
  {Hussey}}, \bibinfo {author} {\bibfnamefont {M.}~\bibnamefont {Abdel-Jawad}},
  \bibinfo {author} {\bibfnamefont {A.}~\bibnamefont {Carrington}}, \bibinfo
  {author} {\bibfnamefont {A.~P.}\ \bibnamefont {Mackenzie}}, \ and\ \bibinfo
  {author} {\bibfnamefont {L.}~\bibnamefont {Balicas}},\ }\href {\doibase
  10.1038/nature01981} {\bibfield  {journal} {\bibinfo  {journal} {Nature}\
  }\textbf {\bibinfo {volume} {425}},\ \bibinfo {pages} {814} (\bibinfo {year}
  {2003})}\BibitemShut {NoStop}%
\bibitem [{\citenamefont {Demsar}\ \emph {et~al.}(1999)\citenamefont {Demsar},
  \citenamefont {Podobnik}, \citenamefont {Kabanov}, \citenamefont {Wolf},\
  and\ \citenamefont {Mihailovic}}]{PhysRevLett.82.4918}%
  \BibitemOpen
  \bibfield  {author} {\bibinfo {author} {\bibfnamefont {J.}~\bibnamefont
  {Demsar}}, \bibinfo {author} {\bibfnamefont {B.}~\bibnamefont {Podobnik}},
  \bibinfo {author} {\bibfnamefont {V.~V.}\ \bibnamefont {Kabanov}}, \bibinfo
  {author} {\bibfnamefont {T.}~\bibnamefont {Wolf}}, \ and\ \bibinfo {author}
  {\bibfnamefont {D.}~\bibnamefont {Mihailovic}},\ }\href {\doibase
  10.1103/PhysRevLett.82.4918} {\bibfield  {journal} {\bibinfo  {journal}
  {Phys. Rev. Lett.}\ }\textbf {\bibinfo {volume} {82}},\ \bibinfo {pages}
  {4918} (\bibinfo {year} {1999})}\BibitemShut {NoStop}%
\bibitem [{\citenamefont {Kusar}\ \emph {et~al.}(2008)\citenamefont {Kusar},
  \citenamefont {Kabanov}, \citenamefont {Demsar}, \citenamefont {Mertelj},
  \citenamefont {Sugai},\ and\ \citenamefont
  {Mihailovic}}]{PhysRevLett.101.227001}%
  \BibitemOpen
  \bibfield  {author} {\bibinfo {author} {\bibfnamefont {P.}~\bibnamefont
  {Kusar}}, \bibinfo {author} {\bibfnamefont {V.~V.}\ \bibnamefont {Kabanov}},
  \bibinfo {author} {\bibfnamefont {J.}~\bibnamefont {Demsar}}, \bibinfo
  {author} {\bibfnamefont {T.}~\bibnamefont {Mertelj}}, \bibinfo {author}
  {\bibfnamefont {S.}~\bibnamefont {Sugai}}, \ and\ \bibinfo {author}
  {\bibfnamefont {D.}~\bibnamefont {Mihailovic}},\ }\href {\doibase
  10.1103/PhysRevLett.101.227001} {\bibfield  {journal} {\bibinfo  {journal}
  {Phys. Rev. Lett.}\ }\textbf {\bibinfo {volume} {101}},\ \bibinfo {pages}
  {227001} (\bibinfo {year} {2008})}\BibitemShut {NoStop}%
\bibitem [{\citenamefont {Stojchevska}\ \emph {et~al.}(2011)\citenamefont
  {Stojchevska}, \citenamefont {Kusar}, \citenamefont {Mertelj}, \citenamefont
  {Kabanov}, \citenamefont {Toda}, \citenamefont {Yao},\ and\ \citenamefont
  {Mihailovic}}]{PhysRevB.84.180507}%
  \BibitemOpen
  \bibfield  {author} {\bibinfo {author} {\bibfnamefont {L.}~\bibnamefont
  {Stojchevska}}, \bibinfo {author} {\bibfnamefont {P.}~\bibnamefont {Kusar}},
  \bibinfo {author} {\bibfnamefont {T.}~\bibnamefont {Mertelj}}, \bibinfo
  {author} {\bibfnamefont {V.~V.}\ \bibnamefont {Kabanov}}, \bibinfo {author}
  {\bibfnamefont {Y.}~\bibnamefont {Toda}}, \bibinfo {author} {\bibfnamefont
  {X.}~\bibnamefont {Yao}}, \ and\ \bibinfo {author} {\bibfnamefont
  {D.}~\bibnamefont {Mihailovic}},\ }\href {\doibase
  10.1103/PhysRevB.84.180507} {\bibfield  {journal} {\bibinfo  {journal} {Phys.
  Rev. B}\ }\textbf {\bibinfo {volume} {84}},\ \bibinfo {pages} {180507}
  (\bibinfo {year} {2011})}\BibitemShut {NoStop}%
\bibitem [{\citenamefont {Xiang}\ \emph {et~al.}(2016)\citenamefont {Xiang},
  \citenamefont {Guo}, \citenamefont {Li}, \citenamefont {Cui}, \citenamefont
  {Qian}, \citenamefont {Hussain}, \citenamefont {Liu}, \citenamefont {Yao},
  \citenamefont {Rao},\ and\ \citenamefont {Zou}}]{Xiang2016}%
  \BibitemOpen
  \bibfield  {author} {\bibinfo {author} {\bibfnamefont {H.}~\bibnamefont
  {Xiang}}, \bibinfo {author} {\bibfnamefont {L.}~\bibnamefont {Guo}}, \bibinfo
  {author} {\bibfnamefont {H.}~\bibnamefont {Li}}, \bibinfo {author}
  {\bibfnamefont {X.}~\bibnamefont {Cui}}, \bibinfo {author} {\bibfnamefont
  {J.}~\bibnamefont {Qian}}, \bibinfo {author} {\bibfnamefont {G.}~\bibnamefont
  {Hussain}}, \bibinfo {author} {\bibfnamefont {Y.}~\bibnamefont {Liu}},
  \bibinfo {author} {\bibfnamefont {X.}~\bibnamefont {Yao}}, \bibinfo {author}
  {\bibfnamefont {Q.}~\bibnamefont {Rao}}, \ and\ \bibinfo {author}
  {\bibfnamefont {Z.~Q.}\ \bibnamefont {Zou}},\ }\href {\doibase
  10.1016/j.scriptamat.2016.01.037} {\bibfield  {journal} {\bibinfo  {journal}
  {Scripta Materialia}\ }\textbf {\bibinfo {volume} {116}},\ \bibinfo {pages}
  {36} (\bibinfo {year} {2016})}\BibitemShut {NoStop}%
\bibitem [{\citenamefont {Liang}\ \emph {et~al.}(2006)\citenamefont {Liang},
  \citenamefont {Bonn},\ and\ \citenamefont {Hardy}}]{Liang2006}%
  \BibitemOpen
  \bibfield  {author} {\bibinfo {author} {\bibfnamefont {R.}~\bibnamefont
  {Liang}}, \bibinfo {author} {\bibfnamefont {D.~A.}\ \bibnamefont {Bonn}}, \
  and\ \bibinfo {author} {\bibfnamefont {W.~N.}\ \bibnamefont {Hardy}},\ }\href
  {\doibase 10.1103/PhysRevB.73.180505} {\bibfield  {journal} {\bibinfo
  {journal} {Phys. Rev. B}\ }\textbf {\bibinfo {volume} {73}},\ \bibinfo
  {pages} {180505(R)} (\bibinfo {year} {2006})}\BibitemShut {NoStop}%
\bibitem [{\citenamefont {Gao}\ \emph {et~al.}(2006)\citenamefont {Gao},
  \citenamefont {Ren}, \citenamefont {Shan}, \citenamefont {Wang},
  \citenamefont {Zhang}, \citenamefont {Zhao}, \citenamefont {Yao},\ and\
  \citenamefont {Wen}}]{PhysRevB.74.020505}%
  \BibitemOpen
  \bibfield  {author} {\bibinfo {author} {\bibfnamefont {H.}~\bibnamefont
  {Gao}}, \bibinfo {author} {\bibfnamefont {C.}~\bibnamefont {Ren}}, \bibinfo
  {author} {\bibfnamefont {L.}~\bibnamefont {Shan}}, \bibinfo {author}
  {\bibfnamefont {Y.}~\bibnamefont {Wang}}, \bibinfo {author} {\bibfnamefont
  {Y.}~\bibnamefont {Zhang}}, \bibinfo {author} {\bibfnamefont
  {S.}~\bibnamefont {Zhao}}, \bibinfo {author} {\bibfnamefont {X.}~\bibnamefont
  {Yao}}, \ and\ \bibinfo {author} {\bibfnamefont {H.-H.}\ \bibnamefont
  {Wen}},\ }\href {\doibase 10.1103/PhysRevB.74.020505} {\bibfield  {journal}
  {\bibinfo  {journal} {Phys. Rev. B}\ }\textbf {\bibinfo {volume} {74}},\
  \bibinfo {pages} {020505} (\bibinfo {year} {2006})}\BibitemShut {NoStop}%
\bibitem [{\citenamefont {Zhang}\ \emph {et~al.}(2017)\citenamefont {Zhang},
  \citenamefont {Wang}, \citenamefont {Dong},\ and\ \citenamefont
  {Wang}}]{Zhang2017}%
  \BibitemOpen
  \bibfield  {author} {\bibinfo {author} {\bibfnamefont {S.~J.}\ \bibnamefont
  {Zhang}}, \bibinfo {author} {\bibfnamefont {Z.~X.}\ \bibnamefont {Wang}},
  \bibinfo {author} {\bibfnamefont {T.}~\bibnamefont {Dong}}, \ and\ \bibinfo
  {author} {\bibfnamefont {N.~L.}\ \bibnamefont {Wang}},\ }\href {\doibase
  10.1007/s11467-017-0716-4} {\bibfield  {journal} {\bibinfo  {journal}
  {Frontiers of Physics}\ }\textbf {\bibinfo {volume} {12}},\ \bibinfo {pages}
  {127802} (\bibinfo {year} {2017})},\ \Eprint
  {http://arxiv.org/abs/1708.01991} {arXiv:1708.01991} \BibitemShut {NoStop}%
\bibitem [{\citenamefont {Lu}\ \emph {et~al.}(2018)\citenamefont {Lu},
  \citenamefont {Li}, \citenamefont {Zhang}, \citenamefont {Hwang},
  \citenamefont {Ofori-Okai},\ and\ \citenamefont {Nelson}}]{Lu2018}%
  \BibitemOpen
  \bibfield  {author} {\bibinfo {author} {\bibfnamefont {J.}~\bibnamefont
  {Lu}}, \bibinfo {author} {\bibfnamefont {X.}~\bibnamefont {Li}}, \bibinfo
  {author} {\bibfnamefont {Y.}~\bibnamefont {Zhang}}, \bibinfo {author}
  {\bibfnamefont {H.~Y.}\ \bibnamefont {Hwang}}, \bibinfo {author}
  {\bibfnamefont {B.~K.}\ \bibnamefont {Ofori-Okai}}, \ and\ \bibinfo {author}
  {\bibfnamefont {K.~A.}\ \bibnamefont {Nelson}},\ }\href {\doibase
  10.1007/s41061-018-0185-4} {\bibfield  {journal} {\bibinfo  {journal} {Topics
  in Current Chemistry}\ }\textbf {\bibinfo {volume} {376}},\ \bibinfo {pages}
  {6} (\bibinfo {year} {2018})}\BibitemShut {NoStop}%
\bibitem [{\citenamefont {Born}\ and\ \citenamefont
  {Wolf}(2013)}]{born2013principles}%
  \BibitemOpen
  \bibfield  {author} {\bibinfo {author} {\bibfnamefont {M.}~\bibnamefont
  {Born}}\ and\ \bibinfo {author} {\bibfnamefont {E.}~\bibnamefont {Wolf}},\
  }\href@noop {} {\emph {\bibinfo {title} {Principles of optics:
  electromagnetic theory of propagation, interference and diffraction of
  light}}}\ (\bibinfo  {publisher} {Elsevier},\ \bibinfo {year}
  {2013})\BibitemShut {NoStop}%
\bibitem [{\citenamefont {Fano}(1961)}]{PhysRev.124.1866}%
  \BibitemOpen
  \bibfield  {author} {\bibinfo {author} {\bibfnamefont {U.}~\bibnamefont
  {Fano}},\ }\href {\doibase 10.1103/PhysRev.124.1866} {\bibfield  {journal}
  {\bibinfo  {journal} {Phys. Rev.}\ }\textbf {\bibinfo {volume} {124}},\
  \bibinfo {pages} {1866} (\bibinfo {year} {1961})}\BibitemShut {NoStop}%
\end{thebibliography}%

\clearpage
\appendix
\maketitle
\renewcommand\thefigure{\Alph{section}\arabic{figure}}
\section{Experimental Method}
\makeatletter
\newcommand{\Rmnum}[1]{\expandafter\@slowromancap\romannumeral #1@}
\makeatother

High-quality single crystals of YBa$_2$Cu$_3$O$_{7-\delta}$ (YBCO) were grown by the topseeded solution growth polythermal method using 3BaO-5CuO solvent with size up to 10 mm$\times$10 mm$\times$6 mm \cite{Xiang2016}. Two small pieces of single crystal were cut from the big crystal with size about 5 mm$\times$3 mm$\times$2.5 mm. The one was annealed in flowing nitrogen at 520$^\circ$ for 3 weeks, and the other in flowing argon-oxygen mixture gas at 680$^\circ$ for 3 weeks. The resulting crystals show sharp superconducting transition temperature near 35 K and 55 K as shown in Fig. 1 (a), which indicates that oxygen content of the samples used for optical measurements are roughly about 6.45 and 6.55\cite{Liang2006,PhysRevB.74.020505}. Those two samples are abbreviated to YBCO6.45 and YBCO6.55, respectively.

The optical reflectance spectra along c-axis from far-infrared (FIR) to ultraviolet region (15 - 40000 \cm) were measured by Fourier transform infrared spectrometers (FTIR) (Bruker 113v and Vertex 80v) using a in-situ gold and aluminum overcoating technique. Limited by the signal-noise ratio (SNR) of FTIR spectrometers at low frequency, the harsh raw experimental data in the Terahertz (THz) regime can hardly be used for the calculation of transient optical constants with the multilayer model presented in Appendix \ref{section:multi}. Figure \ref{Fig:R} shows the raw data at two representative temperatures in black solid lines as examples. A fitting model (presented in Appendix \ref{section:fit}) is used for fitting the reflectivity in THz regime (shaded area in Fig. \ref{Fig:R}) and generating the low-frequency explorations (dashed lines in Fig. \ref{Fig:R}). With those reasonable fittings and extrapolations combining the measured data at higher energy scale (colored thin lines), all optical constants in such a broad range can be obtained through Kramers-Kronig transformation.

The equilibrium and photoexcitation induced change of c-axis reflectivity ranging from 10 to 85 \cm were measured by a time-domain THz spectroscopy system, constructed based on an amplified Ti:sapphire laser system with the pulse duration of 35 fs operating at 1 KHz. NIR/MIR pump beam polarized along c-axis is generated by a two output optical parametric amplifier seeded by same white light continuum (two-output OPA). And the THz probe beam is generated and electro-optic sampling (EOS) by 1-mm-thick ZnTe crystals. The spot sizes of NIR (MIR) pump and probe beam at the sample position are $\sim$1.5 mm ($\sim$1 mm) and 0.63 mm, respectively, which will provide nearly homogeneous excitation. The incident angle of the THz probe beam is 30${}^{\circ}$ with the electric field being  perpendicular to the incident plane, i. e. in transverse electric field configuration.

\label{section:expl}
\begin{figure*}[htbp]
\setcounter{figure}{0}
\includegraphics[width=10cm, trim=0 0 0 0,clip]{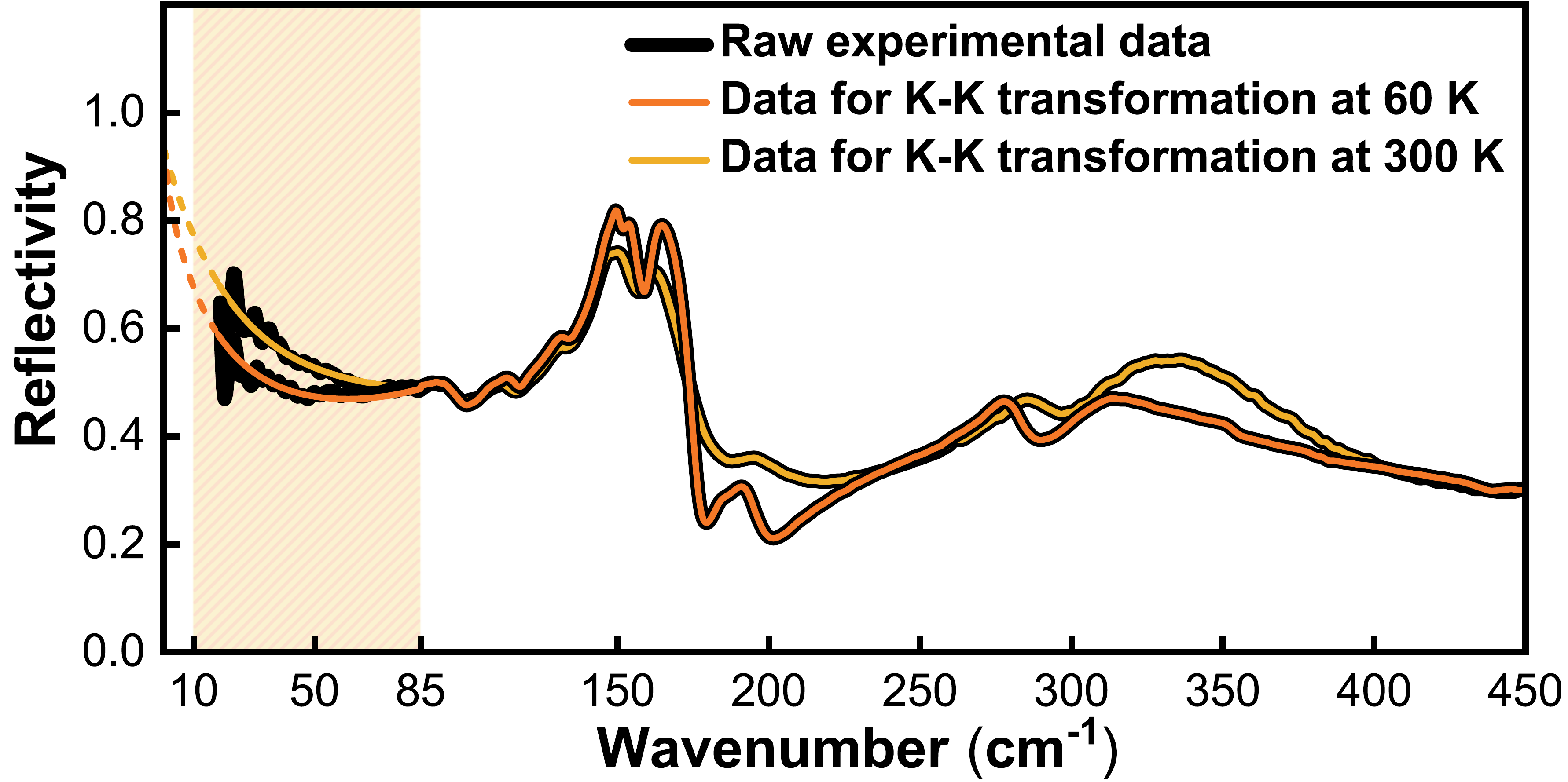}\\
\centering
\caption{Reflectance spectra of YBCO6.55 along c-axis at 60 K and 300 K. The black lines are the raw reflectivity data measured by FTIR spectrometers, and the colored curves are the reflectivity used for Kramers-Kronig transformation.}
\label{Fig:R}
\end{figure*}

In the THz system, two motorized linear translation stages are used to change the relative time delay of optical pump, THz probe and EOS gate beams. One is on the THz probe beam and the other on the EOS gate beam. Time-domain THz profiles at selective delays $\tau$, $\Delta{E}(t, \tau)/E_{peak}$, are measured by moving the stage on THz probe beam. Decay procedure of a specific t of $\Delta{E}(t, \tau)/E_{peak}$ can be measured by moving both those two stages. The decay procedures reported in this work is recorded by fixing t at the maximum position of $\Delta{E}(t, \tau)/E_{peak}$, i.e., $|\Delta{E}_{max}|/E_{peak}$. That measurement configuration assures that the transient responses reflect the authentic light-induced change at a specific time delay $\tau$, and rules out the inaccuracy of relative phase in THz profile. Two choppers are used for modulating the THz probe beam (chopper \Rmnum{1}) and the pump beam (chopper \Rmnum{2}) independently at 377 Hz, which is in favour of  high SNR. Static THz reflected electric field is acquired by using only chopper \Rmnum{1} to modulate the THz probe beam and a lock-in amplifier to read out the balanced EOS diodes. The pump-induced signals are acquired by using only chopper \Rmnum{2} to modulate pump beam and filtering the pump-induced signal with a lock-in amplifier.

The MIR pump pulses with stable carrier envelope phase are generated by difference frequency generation (DFG) with two signal beams from the two-output OPA on a 1 mm thick z-cut GaSe crystal. A low-pass filter is used for blocking the signal pulses after the GaSe crystal. The pulse duration of MIR pump is $\sim$350 fs according to EOS. The MIR pump-induced change disappears when blocking either of the two signal beams used for DFG, which confirms that the MIR pump-induced transient change reported here indeed comes from MIR excitations.

More details about the time-domain THz  experimental setup are presented elsewhere\cite{Zhang2017}.

\section{Determination of static optical reflectance spectrum in low frequency at 5 K}
The optical reflectance spectra in normal state along c-axis of YBCO is almost featureless with only a slight upturn below 100 \cm, e.g. R$_{40 K}(\omega)$ of YBCO6.45, which can by determined by our FTIR spectrometers going down to the lowest measurement frequency 15 \cm directly, as shown in Fig. 1 (a) and Fig. \ref{Fig:1} (a). At 5 K below T$_c$, in order to measure Josephson plasmon edge precisely, a THz time-domain spectrometer is used. The reflected THz electric field ${E}_{5K}(t)$ and ${E}_{40K}(t)$ (Fig. \ref{Fig:1} (b)) are measured by the spectrometer, and the amplitudes of $\tilde{E}_{5K}(\omega)$ and $\tilde{E}_{40K}(\omega)$ can be obtained after doing Fourier transformation (Fig. \ref{Fig:1} (c)). Two different calculation methods are used to maintain the accuracy. The first method uses this equation:
\begin{equation*}
R_{5K}(\omega) = \frac{|\tilde{E}_{5 K}(\omega)|^{2}}{|\tilde{E}_{40 K}(\omega)|^{2}}\cdot R_{40K}(\omega)
\end{equation*}
to determine $R_{5K}(\omega)$. And the second is based on complex reflected coefficient $\tilde{r}(\omega)$:
\begin{equation*}
\begin{split}
& \tilde{r}_{5 K, 30^{\circ}}(\omega) = \frac{\tilde{E}_{5 K, 30^{\circ}}(\omega)}{\tilde{E}_{40 K, 30^{\circ}}(\omega)}\cdot \tilde{r}_{40 K, 30^{\circ}}(\omega), \\
& R_{5K}(\omega)=|\tilde{r}_{5 K,30^{\circ}}(\omega) |^{2}
 \end{split}
\end{equation*}
where $\tilde{r}_{40 K, 30^{\circ}}(\omega)$ is calculated with complex refractive index obtained by FTIR measurements.
In the first method, the determination of $R_{5K}(\omega)$ is not sensitive to phase error of reflected electric field, which may induced by the warming procedure from 5 K to 40 K. In the second one, the incident angel of THz electric field 30${}^{\circ}$ is taken into consideration. $R_{5K}(\omega)$ calculated with those two methods are almost in coincident with each other, as shown in Fig. \ref{Fig:A1}. The optical constants at 5 K is obtained by Kramers-Kronig transformation method, after jointing $R_{5K}(\omega)$ measured by THz spectrometer and FTIR together, in order to avoid the phase sensitivity in THz reflection geometry.

\begin{figure}[htbp]
\setcounter{figure}{0}
\includegraphics[width=5cm, trim=0 0 0 0,clip]{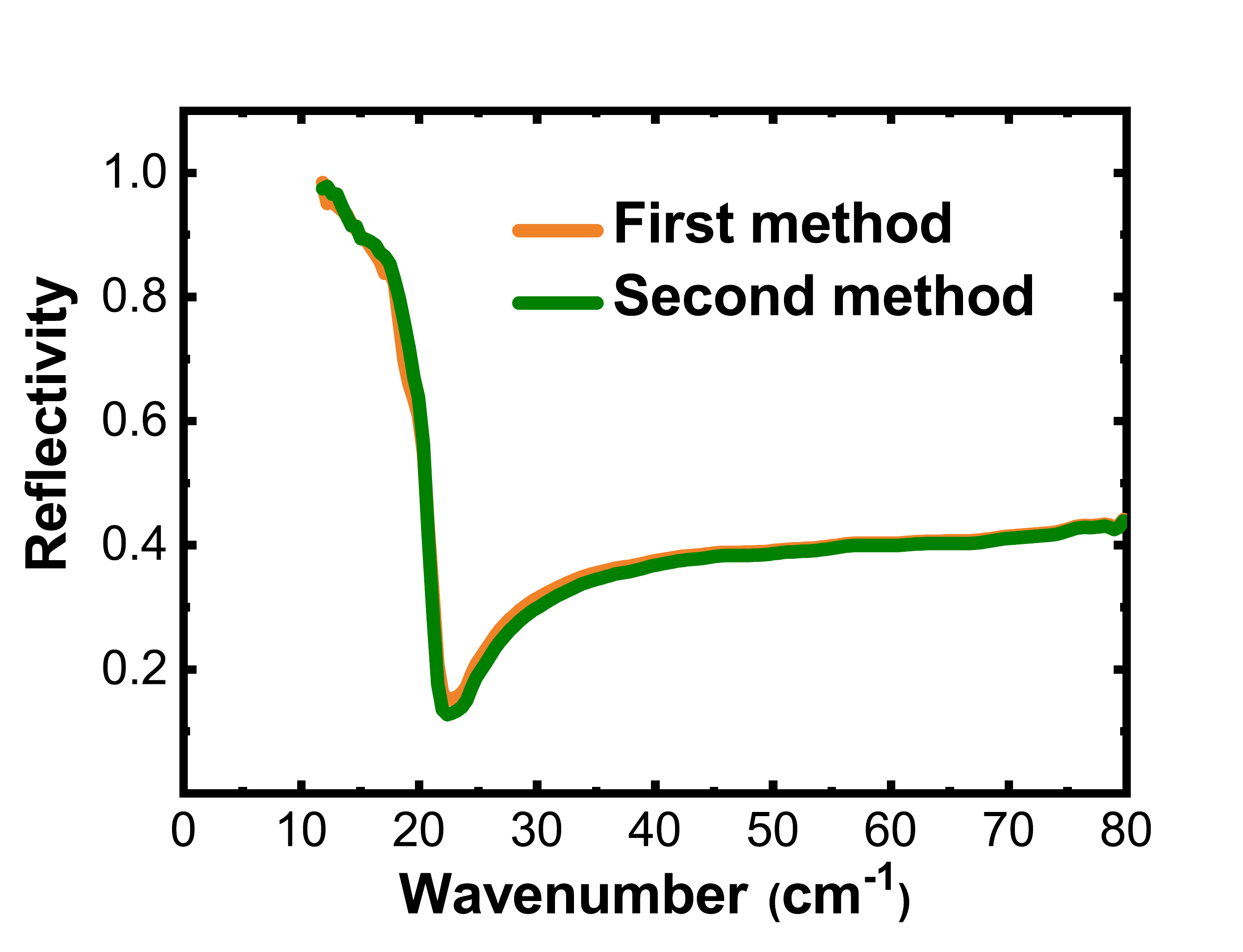}\\
\centering
\caption{Reflectivity at 5 K determined by two different methods.}
\label{Fig:A1}
\end{figure}
\label{section:Rcal}

\section{Determination of the relative phase}
As presented in Appendix \ref{section:expl}, two choppers work independently in the time-domain THz spectroscopy system. So the relative phase between static reflected electric field ${E}_0(t)$ and the pump-induced change at selective time delay $\Delta{E}(t,\tau)$ is unable to know directly. To determine the relative phase, we use the definition of pump-induced change:
$$\Delta{E}(t,\tau) = {E}_{pumped}(t,\tau) -{E}_0(t)$$
where ${E}_{pumped}(t,\tau)$ is measured in the same way as ${E}_0(t)$  but with pump light shedding on the sample.
The relative phase of $\Delta{E}(t,\tau)$ can be determined by that method in superconducting phase, for the pump-induced signal is large enough. But for $\Delta{E}(t,\tau)$ in normal phase, $E_{pumped}(t,\tau) - E_0(t)$ may be disguised by fluctuations at THz peak position.

\begin{figure}[htbp]
\setcounter{figure}{0}
\includegraphics[width=8cm, trim=5 5 5 15,clip]{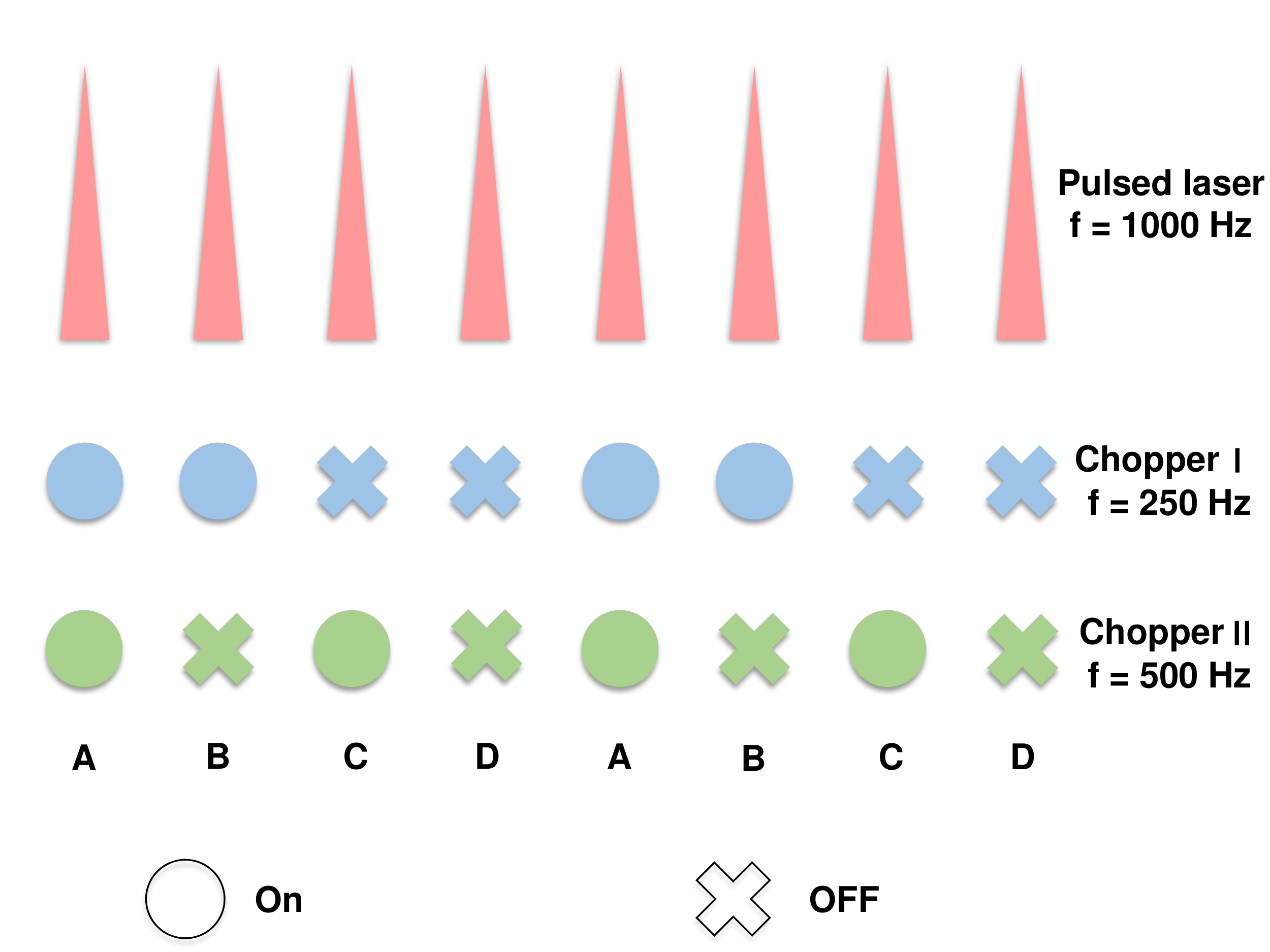}\\
\centering
\caption{Schematic diagram of double modulation technique. "On" represents the light shedding on the sample and "Off" means the light is blocked by the blades of choppers.}
\label{Fig:A3}
\end{figure}
\label{section:phase}

A double modulation technique is further used for determining the relative phase of $\Delta{E}(t,\tau)$ in normal state of YBCO. Two choppers is directly triggered by the Ti:sapphire laser system, whose repetition rate is 1 KHz. Pump beam and THz probe beam are modulated at 250 and 500 Hz respectively by those choppers. A multichannel high-speed data acquisition card (DAQ) is used to read out the EOS balanced detectors \cite{Lu2018}. In that configuration, there will be four different cases (A, B, C, D) if the phase of the two choppers are correctly set, as shown in Fig. \ref{Fig:A3}. Signals can be read out by DAQ only in case A and C, i.e. ${E}_{pumped}(t,\tau)$ and ${E}_0(t)$, every four pulses. To distinguish those two different signals read out by DAQ, the phase between chopper \Rmnum{1} and chopper \Rmnum{2} should be fixed using a reference, on which difference between ${E}_{pumped}(t,\tau)$ and ${E}_0(t)$ is significant enough to be distinguished.

For YBCO at 5 K, the difference between ${E}_{pumped}(t,\tau)$ and ${E}_0(t)$, i.e. $\Delta{E}(t,\tau)$, is over 7\% of $E_{peak}$, which can be resolved by the DAQ detection method even the signal to noise ratio is relatively miserable. According to the results measured by the lock-in method presented in Appendix \ref{section:expl}, the phase of chopper \Rmnum{1} and chopper \Rmnum{2} can be fixed to meet the specific condition illustrated in Fig. \ref{Fig:A3}. Then, the phase of $\Delta{E}(t,\tau)$ in normal state of YBCO can be determined.

\section{Multilayer model for Transient Optical Constants Calculation}
The penetration depth of incident light is estimated with $d$ = $1/{{\omega}{k}}$ in present work, defined as the depth at which the amplitude of electric field inside the material falling to $1/e$  of its original value just beneath the surface. $\omega=2\pi f$, $f$ is in the unit of wavenumber ($cm^{-1}$), and $k$ is the imagine part of complex refractive index obtained by broadband reflectivity spectra of YBCO along c-axis after Kramers-Kronig transformation. The estimated broadband penetration depth on YBCO6.45 is shown in Fig. \ref{Fig:A2} (a). Similar curves can also be obtained on YBCO6.55.

In our pump-probe experiments, two selective wavelength are used to interrogate the pump-induced change of THz regime. MIR pump pulses are tuned to 15 $\upmu$m (667 \cm) and NIR pump pulses to 1.28 $\upmu$m (7810 \cm). The penetration depths of those two pump pulses are 4 $\upmu$m and 1.3 $\upmu$m on YBCO6.45, 3 $\upmu$m and 0.9 $\upmu$m on YBCO6.55. In contrast, the penetration depth in THz regime are widely greater than 20 $\upmu$m and even above. That non-negligible mismatch of the penetration depth of pump and probe pulses results in the reflected probe field containing a mixed response of both pumped and unpumped portions of the compound. In order to disentangle those two portions and to obtain the authentic pump-induced change of optical properties, we use a multilayer model assuming that unpumped region lies beneath the pumped region.

When electromagnetic wave, whose wavelength is $\lambda_{0}$ in vacuum, propagates in a non--magnetic layer with a thickness of $z$, the characteristic matrix $\mathbf{M}(z)$ can be written as
\[\mathbf{M}(z)=\begin{bmatrix} cos(k_{0}\tilde{n}zcos\theta_{0})& -\frac{i}{\tilde{p}}sin(k_{0}\tilde{n}zcos\theta_{0}) \\ -{i}{\tilde{p}}sin(k_{0}\tilde{n}zcos\theta_{0})&cos(k_{0}\tilde{n}zcos\theta_{0}) \end{bmatrix}\quad,\]
where $\theta_{0}$ is the angle of incidence and $k_{0} = 2\pi/\lambda_{0}$. In a stratified medium as a pile of homogeneous thin films, it is assumed that many homogeneous thin layers with evolving $\tilde{n}$ stack together along the direction of propagation $z$. If each layer is thin enough, the characteristic matrix of each layer can be written as
\[\mathbf{M}_{j}=\left[ \begin{array}{cc}{1} & {-\frac{\mathrm{i}}{\tilde{p}_{j}} k_{0} \tilde{n}_{j} \delta z_{j} \cos \theta_{j}} \\ {-\mathrm{i} \tilde{p}_{j} k_{0} \tilde{n}_{j} \delta z_{j} \cos \theta_{j},} & {1}\end{array}\right].\]
The characteristic matrix of the total medium can be written as a product of the matrices for each layer,
\begin{equation}
\mathbf{M}=\prod_{j=1}^{N} \mathbf{M}_{j}=\left[ \begin{array}{cc}{1} & {-\mathrm{i} k_{0} B} \\ {-\mathrm{i} k_{0} A} & {1}\end{array}\right],\label{con:1}
\end{equation}
where
$$
A=\sum_{j=1}^{N} \tilde{p}_{j} \tilde{n}_{j} \delta z_{j} \cos \theta_{j}, B=\sum_{j=1}^{N} \frac{\tilde{n}_{j}}{\tilde{p}_{j}} \delta z_{j} \cos \theta_{j}.
$$
Detailed formula derivations can be found in Ref. \citenum{born2013principles}.

For the transverse electric field configuration case in our experiments, $\tilde{p}_{j}=\tilde{n}_{j}cos\theta_{j}$. In the multilayer model we used, the pumped region is assumed as many homogeneous thin layers stacking together along the direction of propagation, with the pump-induced change of refractive index of each pumped layer evolving in exponential decay, $\tilde{n}(z) = \tilde{n}_{0} + \Delta \tilde{n}\cdot e^{-z/l_{p}}$, where $l_p$ is the penetration depth of pump pulses. The expression for A and B can be rewritten as an integral:
$$
A =-\mathrm{i} k_{0} \int^{l_{p}}_{0} \tilde{n}(z)^{2} \cos ^{2} \theta(z) \mathrm{d} z, B =-\mathrm{i} k_{0} \int^{l_{p}}_{0} \mathrm{d} z
$$
Hence the elements of the characteristic matrix are:
\begin{equation*}
\begin{split}
&m_{11} = m_{22} = 1\\
&m_{12} = -ik_{0}l_{p}\\
&m_{21} = -ik_{0}l_{p}(\tilde{n}_{0}^{2}+2(1-e^{-1})\tilde{n}_{0}\Delta \tilde{n} + \frac{1-e^{-2}}{2}\Delta\tilde{n}^{2}-sin^{2}\theta_{0})
\end{split}
\end{equation*}

\begin{figure*}[htbp]
\setcounter{figure}{0}
\includegraphics[width=10cm, trim=0 0 0 0,clip]{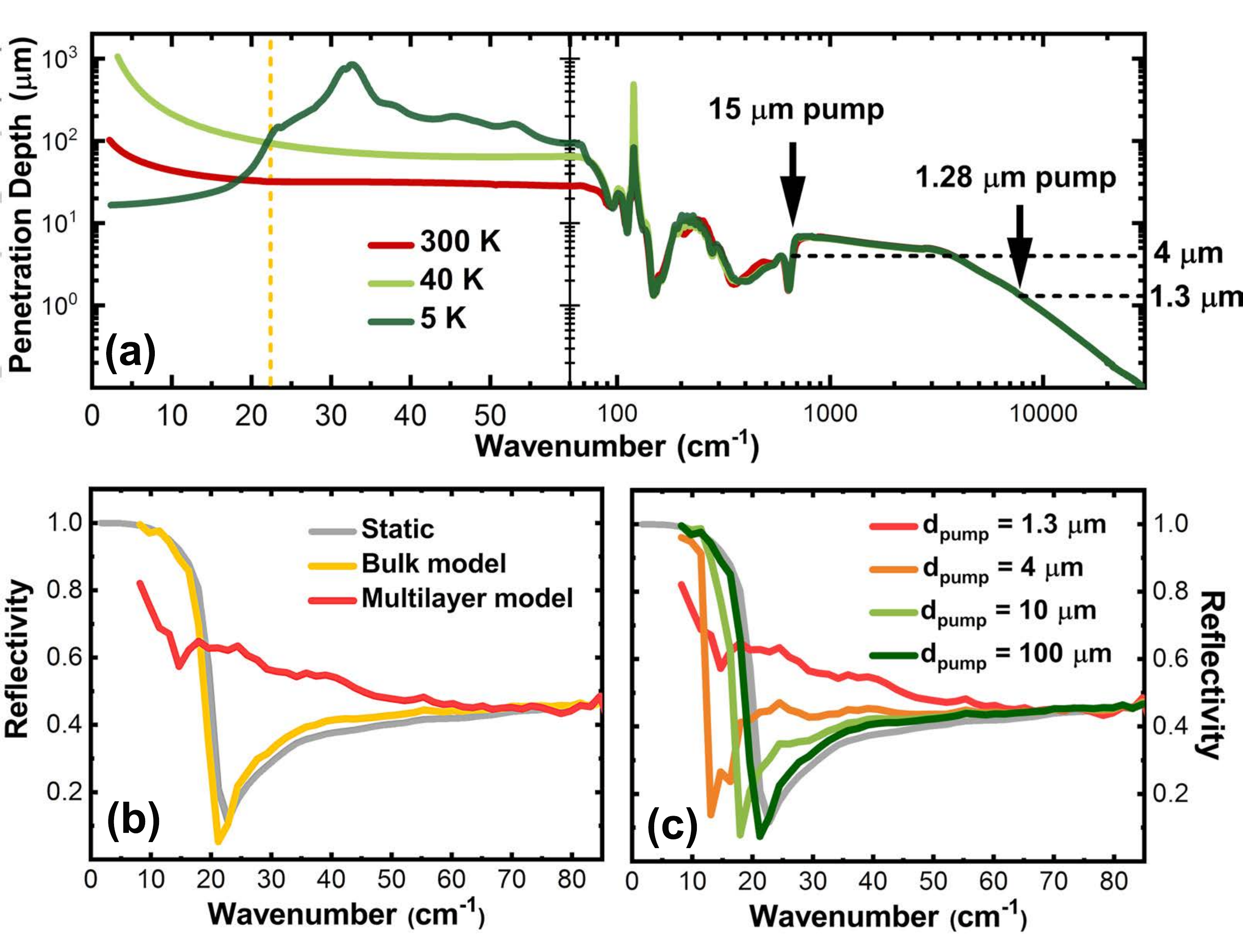}\\
\centering
\caption{A multilayer model should be used to obtain the authentic pump-induced change of optical properties, for the non-neglecting penetration depth difference between pump and probe pulse. (a) the  penetration depth of incident light on YBCO6.45 estimated by the imagine part of complex refractive index. (b) the transient reflectivity calculates with bulk model and multilayer model. (c) The pump-induced change gets weaker when the penetration depth of pump $l_{p}$ gets deeper, and will be in coincident with bulk model if we assume $l_{p}$ is comparable with the penetration depth of probe pulses.}
\label{Fig:A2}
\end{figure*}

We denote the vacuum as medium 1, the pumped region of the material as medium 2, and the unpumped region as medium 3. The reflection coefficient acquired by time-domain THz spectroscopy measurements $\tilde{r}$ can be expressed by
\begin{equation*}
\tilde{r}=\frac{(m_{11}+m_{12}\tilde{p}_{3})\tilde{p}_{1}-(m_{21}+m_{22}\tilde{p}_{3})}{(m_{11}+m_{12}\tilde{p}_{3})\tilde{p}_{1}+(m_{21}+m_{22}\tilde{p}_{3})}
\end{equation*}
where $\tilde{r}_{13}$ is the equilibrium complex reflection coefficient which can be acquired by FRIT measurements and Kramers-Kronig transformation, and $\tilde{r}_{12}$ is the pump-induced transient complex reflection coefficient considering the penetration depth mismatch which is waiting for the following calculation, $\tilde{p}_{1} = cos\theta_{0}$ and $\tilde{p}_{3}=\frac{1-\tilde{r}_{13}}{1+\tilde{r}_{13}}cos\theta_{0}$.
Hence, $\Delta \tilde{n}$ can be solved as an unknown of a quadratic equation shown below
\begin{equation*}
\begin{split}\frac{1-e^{-2}}{2}ik_{0}l_{p}\cdot\Delta \tilde{n}^{2} +2(1-e^{-1})ik_{0}l_{p}\tilde{n}_{0}\cdot\Delta \tilde{n}+(1-ik_{0}l_{p}\tilde{p}_{3})\tilde{p}_{1}\frac{1-\tilde{r}}{1+\tilde{r}}\\
+(ik_{0}l_{p}\tilde{n}_{0}^{2}-ik_{0}l_{p}sin^{2}\theta _{0}-\tilde{p}_{3})=0.
\end{split}
\end{equation*}
There may exist two roots for the quadratic equation according to the quadratic formula. The way to pick a reasonable solution is to maintain the real part of $\tilde{n'}= \tilde{n}_{0} + \Delta \tilde{n}$ positive, for the calculated results should keep in line with the definition of physical quantities.

To check the reasonability of the multilayer model, we now compare the calculated results with different $l_{p}$. If the penetration depth mismatch between pump and probe pulses is neglected, the probed region will be seen as being uniformly pumped, which is defined as ``bulk model". The transient reflectivity can be calculated using the second method in Appendix \ref{section:Rcal}:
\begin{equation*}
\begin{split}
& \tilde{r'}(\omega,\tau) = \frac{\tilde{E'}(\omega,\tau)}{\tilde{E}_{5 K}(\omega)}\cdot \tilde{r}_{5 K}(\omega), \\
& R'(\omega,\tau)=| \tilde{r'}(\omega,\tau)|^{2}
 \end{split}
\end{equation*}
where $\tilde{E'}(\omega,\tau)$ is the Fourier transformation of ${E'}(t,\tau) = {E}_{5 K}(t) +\Delta{E}(t,\tau)$.

Figure \ref{Fig:A2} (b) and (c) compare the transient reflectivity calculated with different models and $l_{p}$, taking the result of the superconducting state of YBCO 6.45 at 9 ps after the excitation of 1.28 $\mu m$ pump by a fluence of 1 mJ/cm$^2$ as an example. Using bulk model or multilayer model, the pump-induced change, i.e. reflectivity being suppressed below the static edge and enhanced at higher energy scale. The pump-induced effects will get weaker if we assume a deeper penetration of pump pulses, as shown in Fig. \ref{Fig:A2} (c). When $l_{p}$ is assumed to be 100 $\upmu m$, which is comparable with the penetration depth of probe, the results calculated using multilayer model seems nearly the same with that using bulk model. That validates the application of the multilayer model we use.
\label{section:multi}

\section{Dataset of Higher Pump Fluence}

Intense ultrafast laser pulses now provide a new route to manipulate the structural and electrical properties of a quantum material. Strong electric field of pump pulses seems to be the key for those manipulations \cite{PhysRevLett.113.026401, RN336, Kaiser2014, PhysRevX.9.021036,liu2019pump}. In the measurements reported here, the pulse durations of pump pulses are 50 fs for NIR pump and 350 fs for MIR pump, which makes it possible to attain sufficient peak electric field without introducing overwhelming laser-induced heating. In the main text, the minimum peak electric field of NIR (MIR) pump is 3.9 MV/cm (1.5 MV/cm) although the fluence is only 1 mJ/cm$^2$.

\begin{figure*}[htbp]
\setcounter{figure}{0}
\includegraphics[width=14cm, trim=0 0 0 0,clip]{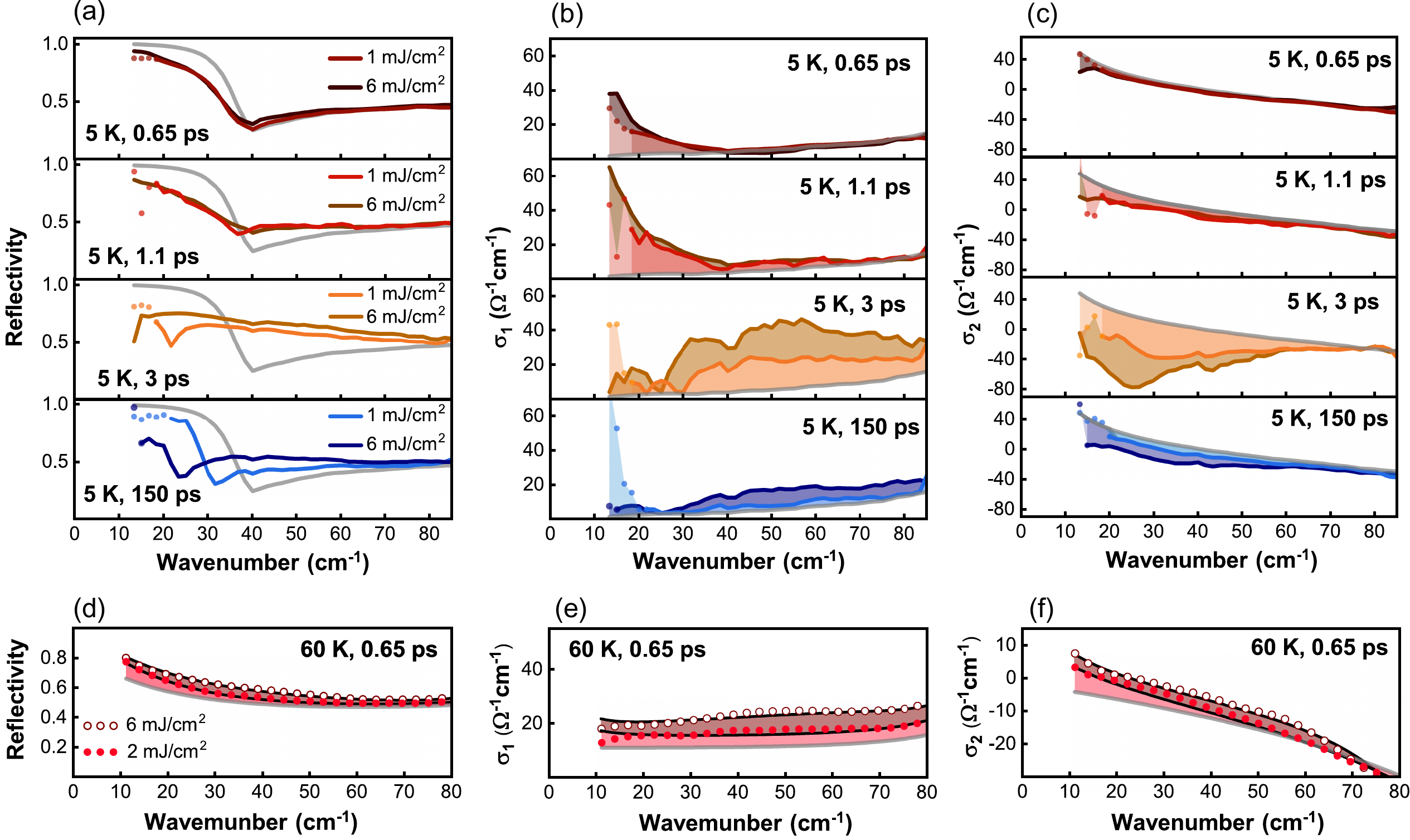}\\
\centering
\caption{ Transient reflectivity $R(\omega, \tau)$, real part of conductivity $\sigma_{1}(\omega, \tau)$ and imaginary part of conductivity $\sigma_{2}(\omega, \tau)$ after being excited by NIR pump pulses at a higher NIR pump fluence of 6 mJ/cm$^2$ with the peak electric field at 9.5 MV/cm in superconducting (upper panel) and normal (lower panel) states.}
\label{Fig:high}
\end{figure*}

A recent manuscript on arXiv presents a pump-induced depletion of superconduting condensate below T$_{c}$ with 19.2 THz pump at a fluence of 1.5 mJ/cm$^2$, and an increase of superfluid density in superconduting state when increase pump fluence to 8 mJ/cm$^2$ with the peak electric field at 3 MV/cm \cite{liu2019pump}. We could not make a comparison on MIR regime due to the limitation on MIR pump fluence in our experiments. Here we repeat our experiments at a higher NIR pump fluence of 6 mJ/cm$^2$ and peak electric field of $\sim$9.5 MV/cm in both supercondcting and normal state as shown in Fig. \ref{Fig:high}, in which the peak electric field is much higher than other reports on YBCO.

In superconducting state, similar but more significant light-induced effects are observed after being excited by higher pump fluence. As increasing the NIR pump fluence, the Josephson plasmon edge in $R(\omega, \tau)$ is suppressed to lower energy scale as shown in Fig. \ref{Fig:high} (a), accompanied by an increase of $\sigma_{1}(\omega, \tau)$ as shown in Fig. \ref{Fig:high} (b), which indicates a more severe depletion of superconductivity and notable quasiparticle excitations. The reduction in $\sigma_{2}(\omega, \tau)$ presented in Fig. \ref{Fig:high} (c) also reveals the depletion of superconductivity.

In normal state, $R(\omega, \tau)$, $\sigma_{1}(\omega, \tau)$ and $\sigma_{2}(\omega, \tau)$ plotted in scatters are enhanced simultaneously once increasing pump fluence, which can still be well reproduced by the Drude-Lorentz model (solid lines) with enhanced plasma frequencies of Drude components. The fitting parameters are presented in Table \ref{tab:transient}.

\label{section:multi}

\section{Fitting Model and Parameters}
\label{section:fit}
A Drude-Lorentz model is used for fitting optical constants in both equilibrium non-equilibrium state. The model are composed by two Drude components and 13 Lorentz ones, which are used for depicting out-of-plane charge conduction, and phonons/pseudogap, respectively,
 \begin{equation}
\epsilon(\omega)=\epsilon_\infty-\sum_{i=1}{{\omega_{pi}^2}\over{\omega^2+i\omega/\tau_i}}+\sum_{j=1}{{S_j^2}{e^{i\theta_{j}}}\over{\omega_j^2-\omega^2-i\omega/\tau_j}}.
\label{chik}
\end{equation}
$\epsilon_\infty$ is taken as 4.5 here for the contribution from ions and high energy electronic excitations. In the Drude terms, $\omega_{pi}$ is plasmon frequency being proportional to $N/m^{*}$, in which $N$ is the density of charge carriers and $m^{*}$ is the effective mass of carriers. In the Lorentz terms, $S_j$ is effective plasmon frequency reflecting the oscillation strengths, $\theta_j$ is the asymmetric parameter for Fano phonons \cite{PhysRev.124.1866} and $\omega_j$ is the center frequency of excitations. $1/\tau_i$ and $1/\tau_j$ is scattering rate of excitations.

The fitting results are plotted in Fig. 1 (c) (d) for equilibrium state, Fig. 3 (d)-(f) and Fig. \ref{Fig:high} for non-equilibrium state. Fitting parameters for Drude components are presented in Table \ref{tab:static} and Table \ref{tab:transient}.

\begin{table}[h]  
\caption{Fitting parameters of equilibrium state}  
\begin{tabular}{ccccc}
\toprule
Temperature (K)& $\omega_{p1} (cm^{-1})$& $1/\tau_1$ ($cm^{-1}$)& $\omega_{p2} (cm^{-1})$& $1/\tau_2$ ($cm^{-1}$)\\
\midrule
60& 56& 60& 420& 270\\
300& 65& 8& 550& 270\\
\bottomrule
\end{tabular}
\label{tab:static}
\end{table}

\begin{table*}[h]  
\caption{Fitting parameters of non-equilibrium state at 60 K $^{\rm *}$}  
\begin{tabular}{cccccc}
\toprule
Pump Fluence (mJ/cm$^2$)&Time Delay (ps)& $\omega_{P1} (cm^{-1})$& $1/\tau_1$ ($cm^{-1}$)& $\omega_{P2} (cm^{-1})$& $1/\tau_2$ ($cm^{-1}$)\\
\midrule
2&0.65& 70& 4& 490& 270\\
2&1.5& 70& 10& 600& 270\\
2&3& 30& 10& 530& 270\\
6&0.65& 85& 4& 520& 270\\
\bottomrule
\end{tabular}\\
\label{tab:transient}
\footnotesize{$^{\rm *}$ may only reflect the trend of free carrier evolution for the reasons mentioned in the main text.}\\
\end{table*}

In the static state, a notably small scattering rate $1/\tau_1$ is used for fitting as increasing temperature, especially at 300 K, which results from the dramatic upturn develops in reflectivity below 90 \cm as shown in Fig. 1 (b). Higher plasma frequencies with adjustable scattering rates are used to reproduce the optical constants of photoexcited state.  As we have mentioned in the discussion part, the charge conduction along c-axis of cuprate superconductors is rather complicated and not capable of being explained by simple Drude models. Although the Drude-Lorentz model is able to reproduce the optical constants in equilibrium and transient states, the fitting parameters should be considered to indicate the trend of evolution.

Apart from that, some defects can result from the measuring range of transient optical constants. In YBCO, the phonons dominate the c-axis dynamics, which is also the main character for parameter fitting. However, the light-induced change at higher energy scale, especially near phonon positions, is unknown for the limitation of THz probe beam generated by ZnTe. As a result, fitting of transient response in THz regime without any information of phonons is somehow tricky. For example, the phonon locates near 95 \cm at 5 K exhibits an asymmetric feature, which indicates a carrier-phonon interaction and the asymmetric parameter $\theta_j$ is non-zero. It can be anticipated that the phonon will also be changed after excitation, for there are still significant divergences between transient and static optical constants up to 80 \cm, as shown in Fig. 3 in main text. The transient response of the phonon turn to expand to THz regime inevitably, which can lead to some inaccuracy in fitting procedure/parameters.

\section{Dataset of YBCO6.45}

We also performed a full set of experiments on YBCO6.45, and the results are similar with that on YBCO6.55 presented in the main text in both superconducting and normal state. Here are all the data collected on YBCO6.45.

\begin{figure*}[htbp]
\setcounter{figure}{0}
\includegraphics[width=18cm, trim=0 0 0 0,clip]{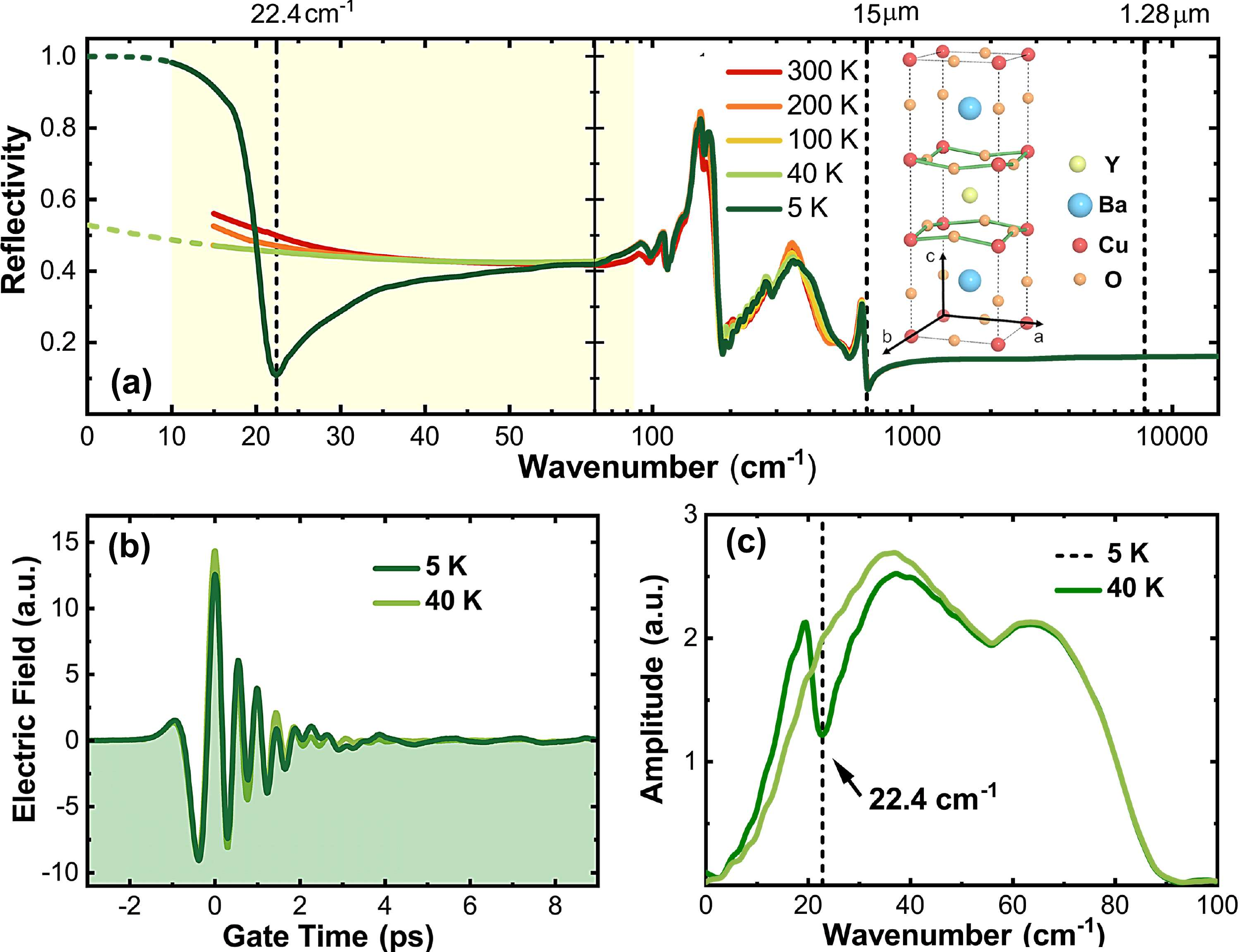}\\
\centering
\caption{ Broadband reflectivity spectra R($\omega$) of YBa$_2$Cu$_3$O$_{6.45}$ along c-axis. (a) The out-of-plane reflectance spectra behave in an insulating manner in normal state and are governed by phonons in FIR region. In superconducting state, Josephson plasmon edge develops near 22.4 \cm. (b) reflected THz electric field ${E}_0(t)$ measured by a time domain THz spectroscopy system. (c) amplitude of Fourier transformation of ${E}_0(t)$, i.e. $|{\tilde{E}}_0(\omega)|$. Reflectivity at 5 K in shaded area of (a) can be calculated with those values as presented in Appendix \ref{section:expl}.}
\label{Fig:1}
\end{figure*}

\begin{figure*}[htbp]
  \centering
\setcounter{figure}{1}
\includegraphics[width=18cm, trim=0 0 0 0,clip]{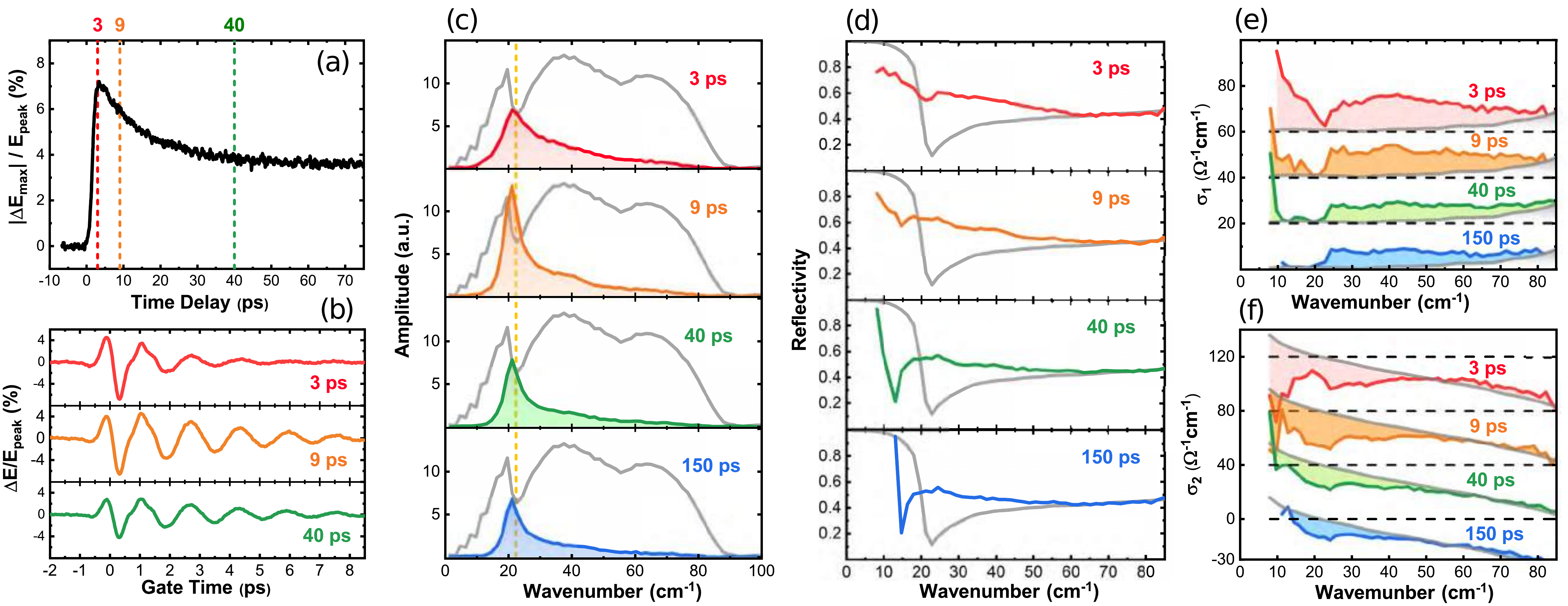}\\
  \caption{Pump-induced changes after excitatiing by 1.28 $\mu m$ with the fluence of 1 mJ/cm$^2$ at 5 K. (a)  the decay procedure of the maximum value of pump-induced change of the reflected THz electric field, $|\Delta E_{max}|/E_{peak}$. $E_{peak}$ is the maximum value of ${E}_0(t)$. (b) the pump induced relative change $\Delta{E}$(t, $\tau$)/$E_{peak}$ in time domain at 3 ps, 9 ps and 40 ps. (c) Fourier transformed spectrum of $\Delta{E}(t, \tau)$. Grey lines is the Fourier transformation of static reflected electric field divided by a coefficient. (d) transient reflectivity $R(\omega, \tau)$ (e) real part of conductivity, $\sigma_{1}(\omega, \tau)$ (f) imaginary part of conductivity, $\sigma_{2}(\omega, \tau)$. Optical constants in static state are plotted in grey lines. In superconducting state, the superconducting condensate of YBCO6.45 is heavily disturbed or destroyed upon NIR pumping and gradually recovered with time delays, which is exactly the same with that on YBCO6.55.}\label{Fig:DeltaE}
\end{figure*}

\begin{figure*}[htbp]
\setcounter{figure}{2}
\includegraphics[width=16cm, trim=0 0 0 0,clip]{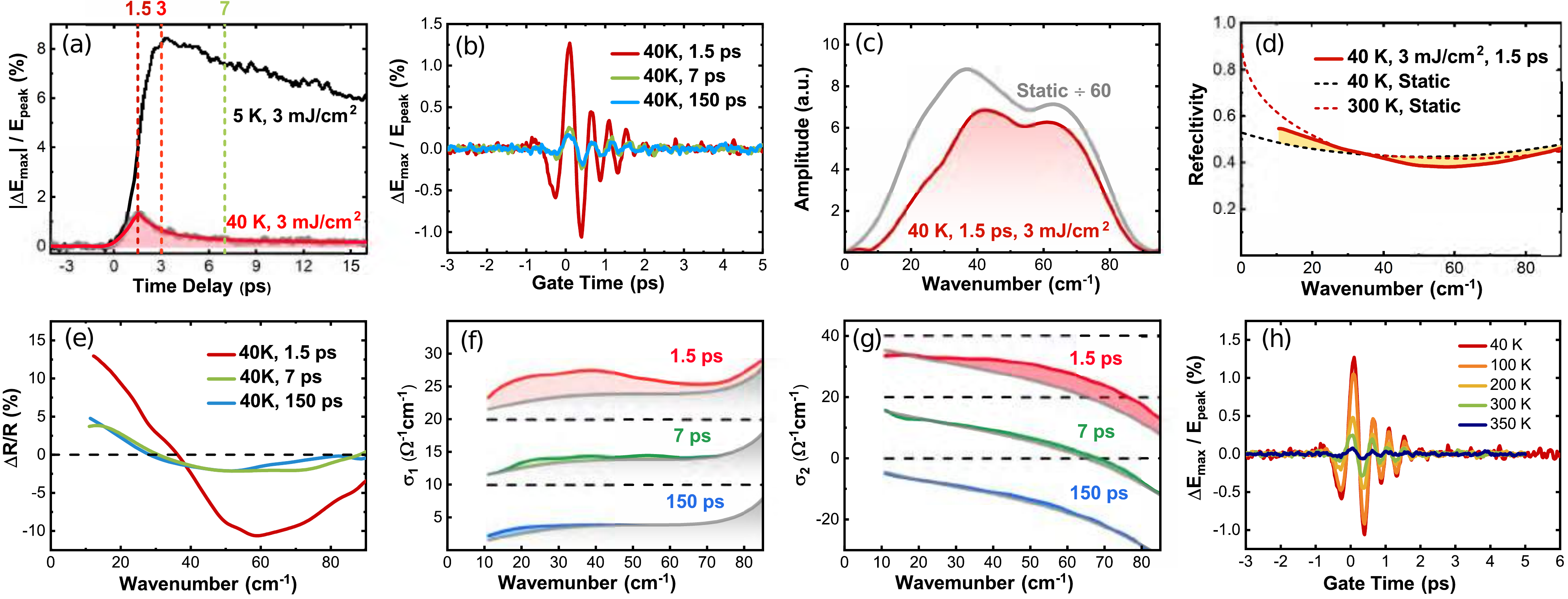}\\
  \caption{Pump-induced changes after excitation at 1.28 $\mu m$ by a fluence of 3 mJ/cm$^2$ in normal state. (a)   the decay procedure of the maximum value of pump-induced change of the reflected THz electric field, $|\Delta E_{max}|/E_{peak}$, at 40 K (experimental results are ploted in grey dots, and the red line is guide for the eyes) and 5 K (black line). (b) the pump-induced relative change $\Delta{E}$(t, $\tau$)/$E_{peak}$ in time domain at three different time delays. (c) the Fourier transformed spectrum of $\Delta{E}(t, 1.5 ps)$. Grey lines is the Fourier transformation of static reflected electric field divided by a coefficient, 60. (d) the transient reflectivity at 1.5 ps. Dashed lines are the static reflectivity at 5 K and 300 K for comparison. (e) there is a very weak edge-like structure with a minimum (the starting point of upturn) near 60 \cm can be observed in the relative change of transient reflectivity at 40 K, which is slightly different with that on YBCO6.55 with only enhancement of reflectivity being observed. It is widely known that there are more free carriers in higher doping levels, which results in a higher plasmon edge in reflectivity as shown in Fig. 1 (b). So it can be concluded that the starting point of upturn on YBCO6.55 may be at higher energy scale and out of the measuring range. (f) transient real part of conductivity $\sigma_{1}(\omega, \tau)$ (g) imaginary part of conductivity $\sigma_{2}(\omega, \tau)$ after excitations. Optical constants in static state are plotted in grey lines. (h) the temperature dependence of pump induced relative change $\Delta {E}$(t, $\tau$)/$E_{peak}$ shows that those transient response sustains above room temperature. }\label{Fig:3}
\end{figure*}

\begin{figure*}[htbp]
  \centering
  \setcounter{figure}{3}
\includegraphics[width=18cm, trim=0 0 0 0,clip]{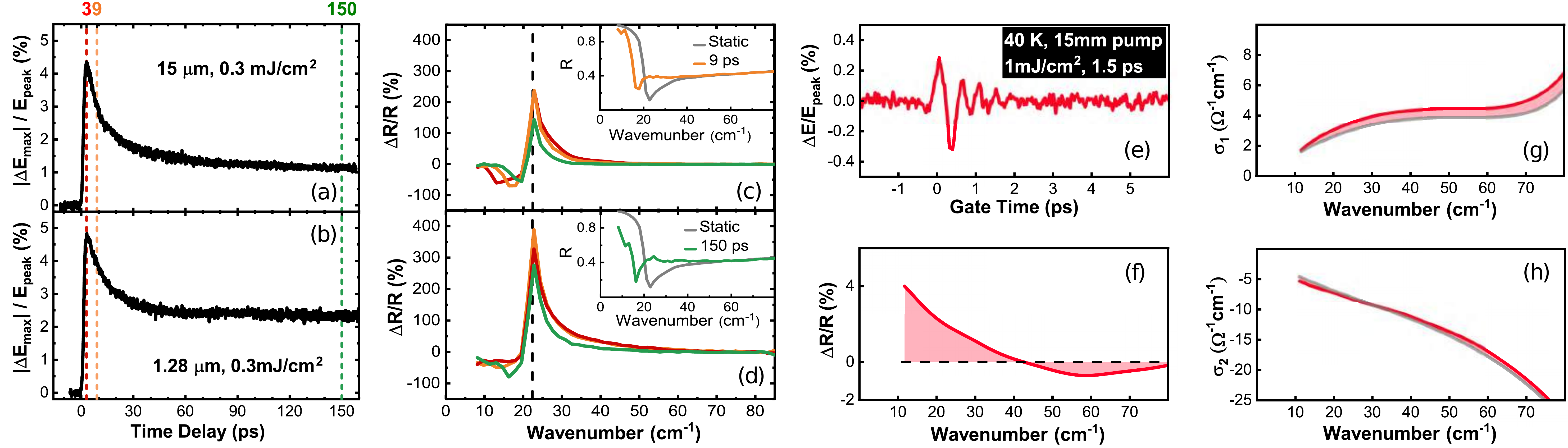}\\
  \caption{ Comparison between the pump-induced effects after the excitation of 15 $\upmu$m and 1.28 $\upmu$m pulses at 0.3 mJ/cm$^2$ (1 mJ/cm$^2$) at 5 K (40 K). (a) and (b) the decay procedure of the maximum value of pump-induced change of the reflected THz electric field, $|\Delta E_{max}|/E_{peak}$ at 5 K and 40 K after the excitation of 15 $\upmu$m and 1.28 $\upmu$m pulses. (c)-(d) the pump-induced relative change of reflectivity at 5 K. Similar to NIR pump pulses, MIR pulses also turn to remove Josephson plasmon edge upon exciting and then drive YBCO into a state with the edge at lower energy scale together with some spectral weight from excited quasiparticles in the superconducting state. (e) the pump-induced change of THz electric field at 40 K after exciting by 15 $\mu m$ pump pulses at the fluence of 1 mJ/cm$^2$ in time domain. (f)-(h) the transient optical constants after excitations by MIR pump pulses at 40 K in the normal state. Grey lines are the optical constants in the equilibrium state. An edge-like shape and enhancement of conductivity can be seen as the clue for quasiparticle excitations, which is qualitatively the same with that after NIR excitations  }\label{Fig:pump15mum}
\end{figure*}

\end{document}